\documentclass[onecolumn,showpacs,amsmath,amssymb,pre,superscriptaddress,a4paper,floatfix,showkeys]{revtex4}
\usepackage{dcolumn}
\usepackage{graphicx}
\usepackage{epsfig}
\usepackage{ulem} 

\newcommand\concentration{n}
\newcommand\tauR{T_R}
\newcommand\tauRR{T_{R0}}

\begin{document}

\title{Quantum simulations and experiments on Rabi oscillations
\\of spin qubits: intrinsic {\sl vs} extrinsic damping\footnote{Accepted for publication in Physical Review B}}%

\author{Hans De Raedt}
\affiliation{%
Department of Applied Physics,
Zernike Institute for Advanced Materials,
University of Groningen, Nijenborgh 4, NL-9747 AG Groningen, The Netherlands
}%
\author{Bernard Barbara}
\affiliation{%
Institut N\'eel, CNRS,  Universit\'e Joseph Fourier, BP 166, F-38042 Grenoble Cedex 9, France
}%
\affiliation{%
Laboratoire de Chimie Inorganique et Biologique (UMR-E3 CEA-UJF),
INAC, CEA-Grenoble,
17 Ave. des Martyrs,
38054 Grenoble Cedex 9,
France
}
\author{Seiji Miyashita}
\affiliation{%
Department of Physics, Graduate School of Science,
The University of Tokyo, 7-3-1 Hongo, Bunkyo-Ku, Tokyo 113-8656
}
\affiliation{%
CREST, JST, 4-1-8 Honcho Kawaguchi, Saitama 332-0012, Japan\\
}%
\author{Kristel Michielsen}
\affiliation{%
Institute for Advanced Simulation, J\"ulich Supercomputing Centre,
Research Centre J\"ulich, D-52425 J\"ulich, Germany
}%
\author{Sylvain Bertaina}
\affiliation{%
IM2NP-CNRS (UMR 6242) and Universit\'{e} Aix-Marseille, Facult\'{e}
des Sciences et Techniques, Avenue Escadrille Normandie Niemen - Case
142, F-13397 Marseille Cedex, France.
}%
\author{Serge Gambarelli}
\affiliation{%
Laboratoire de Chimie Inorganique et Biologique (UMR-E3 CEA-UJF),
INAC, CEA-Grenoble,
17 Ave. des Martyrs,
38054 Grenoble Cedex 9,
France
}
\date{\today}

\begin{abstract}
Electron Paramagnetic Resonance experiments show that the decay of Rabi oscillations of ensembles of spin qubits depends
noticeably on the microwave power and more precisely on the Rabi frequency, an effect recently called ``driven decoherence''.
By direct numerical solution of the time-dependent Schr\"odinger equation of the associated many-body system, we scrutinize the
different mechanisms that may lead to this type of decoherence.
Assuming the effects of dissipation to be negligible ($T_1=\infty$),
it is shown that a system of dipolar-coupled spins with -- even weak-- random inhomogeneities is sufficient
to explain the salient features of the experimental observations.
Some experimental examples are given to illustrate the potential of the numerical simulation approach.
\end{abstract}

\maketitle

\section{Introduction}\label{intro}

Decoherence generally occurs when the phase angle associated with a periodic motion is lost due to some interaction with
exterior noise. In classical mechanics it may apply to classical waves such as sound waves, seismic waves, sea waves, whereas
in quantum mechanics it applies to the phase angles between the different components of a system in quantum superposition.
The loss of phase of a quantum system may bring it to its classical regime, raising the question of whether and how
the classical world may emerge from quantum mechanics.
Together with the claim that decoherence is also relevant to a variety of
other questions ranging from the measurement problem to the arrow of time,
this underlines the important role of decoherence in the foundations of quantum mechanics.
It is for all these reasons that the analysis of decoherence in quantum systems must make
allowance and in particular must distinguish between decoherence induced by the imperfections of real systems and intrinsic
decoherence induced by identified or hidden couplings to the environment.
The different sources of decoherence can be classified in two main categories~\cite{Morello2006},
the one-qubit decoherence coming from the coupling of individual qubits with the environment~\cite{Leggett1987,Weiss1999,PROK00} and
the multi-qubit or pairwise decoherence coming from multiple interactions between
pairs of qubits~\cite{Alicki2002,Terhal2005,Klesse2005,Novias2006}

In this paper we take the example of paramagnetic spins because of the quality of the systems which can be elaborated (single-crystals)
and the possibility, offered by magnetism, to start calculations from first principles.
Here, the one-qubit decoherence is, in general, associated with phonons and hyperfine
couplings~\cite{Abragam1961,Villain1994,Wurger1998,Leuenberger2000}
which are intrinsic effects,
but also with non-intrinsic effects resulting from weak disorder always present in real systems of finite size:
inhomogeneous fields, $g$-factor distributions, and positional distributions.
Multiple-qubit decoherence is generally due to pairwise dipolar interactions
with distant electronic or nuclear qubits, which is an  intrinsic mechanism~\cite{Prokof'ev1996}.
Below, we shall see that, more generally, when pairwise decoherence takes place in the rotating  frame,
extrinsic decoherence becomes crucial by itself and also by amplifying intrinsic decoherence.
In particular, by way of some examples, it will be shown that the origin of driven
decoherence is of the one-qubit type i.e. with multiple possible origins
(depending on the nature disorder).
Even if dominant sources of decoherence may sometimes be identified,
the complete description of decoherence and in particular, the discrimination of
intrinsic and extrinsic decoherence are generally not accessible to experimentalists.
This is a major obstacle for the reduction of decoherence, and it holds beyond magnetism.
We believe that the present, pragmatic approach, should be of great help in common situations
where intrinsic and extrinsic decoherence mechanisms interoperate.

Assuming that each type of decoherence has its own ``signature'' on the Rabi oscillations,
we have started a systematic study in which the Rabi oscillations of an ensemble of
spins are simulated by direct numerical solution of the time-dependent Schr\"odinger equation (TDSE) of the
associated many-body system.
These simulations are performed using a parallel algorithm implementation based on a
massively parallel quantum computer simulator~\cite{RAED07x}.
The various mechanisms that may lead to decoherence of Rabi oscillations are successively implemented in Hamiltonians, leading to different
types of damping, oscillation shapes, non-zero oscillation averages and their evolutions with exterior parameters such as the
microwave power and the applied static field.
The comparison with measured Rabi oscillations allows us to scrutinize the different
decoherence mechanisms and to understand more basic aspects of decoherence,
thereby opening a route to search for the optimal -- intrinsic and extrinsic -- ways to improve coherence
of Rabi oscillations, i.e. the number of oscillations which is important for all applications.

The present study is limited to the decoherence of Rabi oscillations,
that is the decoherence measured immediately after the application of a long microwave pulse.
Following an earlier suggestion~\cite{BOSC93,SHAK97,AGNE99},
it was shown that the microwave pulse inducing Rabi oscillations is itself an important source of
decoherence in all the investigated systems
(``driven decoherence''~\cite{BERT07,BERT08,BERT09a,BERT09b}), 
except when the microwave power is very small, in which case the Rabi frequency is also very small.
As a consequence the number of Rabi oscillations remains nearly constant,
that is one cannot increase it by increasing the microwave power.

This observation can be quantified by comparing
the damping time of Rabi oscillations (Rabi decay time $T_R$) with the usual spin-spin relaxation time $T_2$.
The theoretical results given in this paper are all exact.
Depending on the Hamiltonian parameters, the results were obtained analytically (in simple cases)
and numerically (in more general cases, including dipolar interactions) and covered the large range
of possibilities, namely from $T_R \ll T_2$ upto $T_R \approx 2 T_2$
when dipolar interactions dominate (in the absence of disorder).

The systems used to compare the simulations results with
experimental data are insulating single-crystals of CaWO$_4$:Er$^{3+}$,
MgO:Mn$^{2+}$, and BDPA ($\alpha-\gamma$-bisdiphenylene-$\beta$-phenylally),
a free radical system often used in Electron Paramagnetic Resonance (EPR) calibration.
The latter is not a diluted system, contrary to the two others, but an antiferromagnetic single crystal (identical
environments) with a N\'eel temperature much smaller than the temperature at which our measurements are made (between 4K and 300K).
These systems have been chosen in particular for the differences in their homogeneous/inhomogeneous EPR linewidths.
Furthermore, in these systems the relaxation time $T_1$ is much larger than $T_2$, as this is often the case in solid state systems.
For instance, our experiments yield a $T_1$ which is 10 and 40 times larger than the $T_2$ for MgO:Mn$^{2+}$ and
CaWO$_4$:Er$^{3+}$, respectively. Therefore, as a first step in the theoretical modeling
of these experiments, it is reasonable to neglect the effect of dissipation and focus on the decoherence only.

\begin{figure}[t]
\includegraphics[width=\columnwidth]{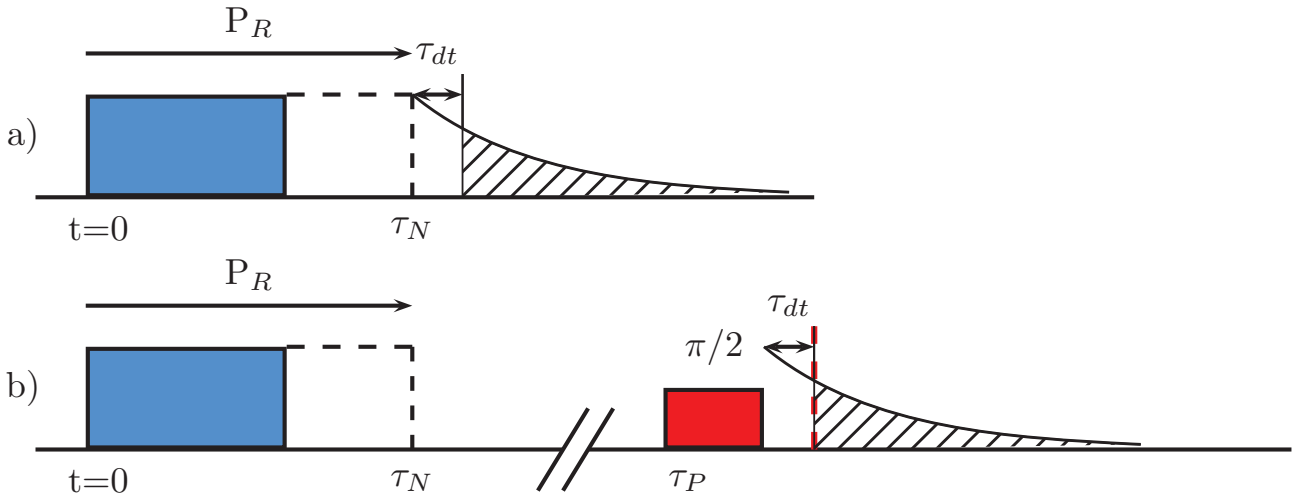}
\caption{(Color online) Pulse sequence used for Rabi oscillation measurements.
a) At $t=0$, a microwave pulse $P_R$ coherently drives the magnetization.
At the end of the pulse ($\tau=\tau_N$) the magnetization component $M^y(t)$ is recorded.
b) After the $P_R$ pulse, one waits a time much
longer than $T_2$ but smaller than $T_1$ such that $M^y(t)$ has vanished.
Then, after the waiting time $T$, a standard Hahn echo sequence of duration $\tau_P$
is used to measure the magnetization component $M^z(t)$.
The dead time of the spectrometer is denoted by $\tau_{dt}$.
}
\label{fig:seq1}
\end{figure}

Rabi oscillations measurements have been performed in a Bruker Elexsys 680 pulse EPR spectrometer working at about $f=9.6$~GHz ($X$-band).
Depending on the sample, measurements have been done at room temperature down to liquid helium temperature (4K).
The static magnetic field has always been chosen to correspond to the middle of the EPR line.
The experimental procedure is illustrated in Fig.~\ref{fig:seq1}.
A microwave pulse $P_R$ starts at $t=0$ and coherently drives the magnetization.
At the end of the pulse ($\tau=\tau_N$) the magnetization is recorded.
Because of the dead time ($\tau_{dt}$) of the spectrometer (about 80ns), it is impossible to directly measure the magnetization right
after the pulse $P_R$.
In this paper, we used two methods for the detection.
The first one is simply to record the free induction decay
(FID) emitted by the system when the microwave field is shut down.
This method gives the value of the magnetization component $M^y(t)$
at the end of the pulse if we take into account two important conditions:
i) $P_R$ is a non selective pulse (all spins of the line are excited)
and under this condition, the FID signal is the Fourier transform of the EPR line.
ii) The EPR linewidth must be sharp enough.
Since the FID is the Fourier transform of the EPR spectrum, a linewidth $\gtrsim 4G$
will lead to a decay time of the FID less than $80$~ns and
the FID will be hidden by the dead time of the spectrometer.
The second method is used when the EPR line is too broad or if one wants to probe the longitudinal
magnetization $M^z(t)$. In this case another probe sequence has to be used.
After the $P_R$ pulse, one waits a time $T$ much
longer than $T_2$ but smaller than $T_1$ in order such that $M^y(t)$ vanishes.
After the waiting time $T$, a standard Hahn echo sequence ($\pi/2-\tau-\pi-\tau-$echo)
is used to measure the longitudinal magnetization.
In the present paper we do not study
(a) the effects of the spin-echo pulses on the measurements
and
(b) the effect of temperature.
For (a), this implies that the comparison with theory is through the measured so-called free-decay time $T_2^\ast$ (different from
the usual $T_2$) in which a component of the total magnetization is directly measured through an induction coil,
and for
(b) that the measurements of $T_2^\ast$ are done at a sufficiently low temperature
which is quite easy to realize since the $T_2^\ast$ of BDPA is nearly independent of temperature,
and more generally the Rabi time $T_R$, most important in the context of this paper,  also.

The paper is organized as follows.
In Section~\ref{model}, the quantum spin model is specified in detail and the simulation procedure is briefly discussed.
Our results are presented in Section~\ref{results}.
As there are many different cases to consider, to structure the presentation,
the results have been grouped according to the kind of randomness,
describing for each kind (i) the non-interacting case, (ii) the interacting case and
(iii) a comparison with experiments if this is possible.
In Section~\ref{BE}, we present a model of ``averaged local Bloch equations'',
giving a complete, exact description of one-qubit decoherence
and incorporates multi-qubit decoherence phenomenologically.
A summary and outlook is given in Section~\ref{conclusions}.

\section{Model}\label{model}

We consider a system of $L$ dipolar-coupled spins subject to a static magnetic field
along the $z$-axis and a circular polarized microwave perpendicular to the $z$-axis.
The Hamiltonian reads
\begin{eqnarray}
H=&-&\mu_B \sum_{j=1}^L \mathbf{B}_j(t)\cdot\mathbf{g}_j\cdot\mathbf{S}_j
+
\frac{\mu_0\mu_B^2}{4\pi} \sum_{j<k}
\frac{\mathbf{S}_j\cdot\mathbf{g}_j\cdot\mathbf{g}_k\cdot\mathbf{S}_k}{\mathbf{r}_{jk}^3}
-3
\frac{(\mathbf{S}_j\cdot\mathbf{g}_j\cdot\mathbf{r}_{jk})(\mathbf{S}_k\cdot\mathbf{g}_k\cdot\mathbf{r}_{jk})}{\mathbf{r}_{jk}^5}
,
\label{B1}
\end{eqnarray}
where $\mathbf{B}_j(t)=(B'_j \cos\omega t, -B'_j \sin\omega t,B_0)$
denotes the external magnetic field, composed of a large static field $B_0$ along the $z$-axis and a
circular time-dependent microwave field $B'_j$ ($\max_j |B'_j|\ll \min_j|B_0|$)
which may depend on the position of the $j$th spin, represented by the spin-1/2 operators
$\mathbf{S}_j=(S_j^x,S_j^y,S_j^z)$ with eigenvalues $\pm1/2$.
The vector $\mathbf{r}_{jk}$ connects the positions of spins $j$ and $k$.
It is assumed that the $g$-tensor can be written as $\mathbf{g}_j=g_e(\openone+\Delta\mathbf{g}_j)$
where the perturbation $\Delta\mathbf{g}_j$ is a random matrix.

As usual in the theory of NMR/ESR, we separate the fast rotational motion induced by the large static field $B_0$
from the remaining slow motion by a transformation to the reference frame that rotates with an angular frequency
determined by $B_0$.
Taking the ideal, non-interacting system without fluctuations in the $g$-tensors as the reference system,
we define $\omega_0=g_e\mu_B B_0$ and assume from now on that this ideal system is at resonance, that is
the microwave frequency is given by $\omega=\omega_0$.

The transformation to the reference frame rotating with angular frequency $\omega_0$ is defined by
\begin{eqnarray}
X_{RF}
&=& \exp\left(it\omega_0\sum_{j=1}^L S^z_j\right) X \exp\left(-it\omega_0\sum_{j=1}^L S^z_j\right)
,
\label{B2}
\end{eqnarray}
where $X$ denotes any combination of spin operators.
Transforming Eq.~(\ref{B1}) to the rotating frame,
we find that $H_{RF}$ contains contributions
that (i) do not depend on time,
(ii) have factors $e^{it\omega_0}$ or $e^{-it\omega_0}$, or
(iii) have factors $e^{2it\omega_0}$ or $e^{-2it\omega_0}$.
Contributions that depend on time oscillate very fast (because $\omega_0$ is large)
and, according to standard NMR/ESR theory, may be neglected, which we have confirmed for a few cases.
The remaining time-independent, secular terms yield the Hamiltonian
\begin{eqnarray}
H_{RF}&=&-\omega_0\sum_{j=1}^L \frac{g^z_j-g_e}{g_e} S^z_j - \mu_B g_e\sum_{j=1}^L B'_j\frac{g^x_j+g^y_j}{2g_e} S^x_j
\nonumber\\
&&+\frac{\mu_0\mu_B^2g_e^2}{4\pi} \sum_{j<k}
\frac{g^z_j g^z_k[1-3z_{jk}^2/r_{jk}^2]}{g_e^2r_{jk}^3} S^z_j S^z_k
+\frac{g^x_j g^x_k[1-3x_{jk}^2/r_{jk}^2]+g^y_j g^y_k[1-3y_{jk}^2/r_{jk}^2]}{2g_e^2r_{jk}^3}
\left(S^x_j S^x_k + S^y_j S^y_k\right)
.
\label{B3}
\end{eqnarray}
From Eq.~(\ref{B3}), it is clear that variations in $g^z_j$ have the same effect as
local variations (inhomogeneities) in the static magnetic field.
Inhomogeneities in the microwave field and the variations in $(g^x_j,g^y_j)$ are cumulative.
Although the variations in $g^x_j$, $g^y_j$, and $g^z_j$ also affect the dipolar interactions,
these effects may be difficult to distinguish from the effect of the positional disorder
of the spins in the solid, in particular if the spins are distributed randomly.
Note that the total magnetization $M^z=\sum_{j=1}^L S^z_j$ commutes
with the dipolar terms in Eq.~(\ref{B3}).
Therefore, neglecting the terms that oscillate with $\omega_0$ or $2\omega_0$,
the longitudinal magnetization $M^z(t)=\sum_{j=1}^L S_j^z$ is a constant of motion
in the absence of a microwave field ($B'_j=0$).

If all the $g$'s are the same and equal to $g_e$ the Hamiltonian Eq.~(\ref{B3}) reduces to the familiar expression
\begin{eqnarray}
H_{RF}&=& - \sum_{j=1}^L h_j S^x_j
+\frac{\mu_0\mu_B^2g_e^2}{4\pi} \sum_{j<k}
\frac{[1-3z_{jk}^2/r_{jk}^2]}{r_{jk}^3} \left[ S^z_j S^z_k-\frac{1}{2}
\left(S^x_j S^x_k + S^y_j S^y_k\right)\right]
,
\label{B4}
\end{eqnarray}
of the Hamiltonian of dipolar-coupled spins in the reference frame that rotates at the resonance frequency $\omega_0$.

\subsection{Simulation model}\label{modelparameter}
We now specify the model as it will be used in our computer simulations.
We rewrite the Hamiltonian Eq.~(\ref{B3}) as
\begin{eqnarray}
H_{RF}/\hbar&=&
-2\pi F_0\sum_{j=1}^L \xi^z_j S^z_j - 2\pi h_p F_{\mathrm{R}}\sum_{j=1}^L \frac{(1+\zeta_j)(2+\xi^x_j+\xi^y_j)}{2} S^x_j
\nonumber\\
&&
+2\pi D_0 \sum_{j<k} \frac{(1+\xi^z_j)(1+\xi^z_k) [1-3z_{jk}^2/r_{jk}^2]}{r_{jk}^3} S^z_j S^z_k
\nonumber\\
&&
+2\pi D_0 \sum_{j<k}
\frac{(1+\xi^x_j)(1+\xi^x_k)[1-3x_{jk}^2/r_{jk}^2]+(1+\xi^y_j)(1+\xi^y_k)[1-3y_{jk}^2/r_{jk}^2]}{2r_{jk}^3}
\left(S^x_j S^x_k + S^y_j S^y_k\right)
,
\label{M0}
\end{eqnarray}
where we take $F_0=\omega_0/2\pi\hbar=9.7\,\mathrm{GHz}$ for the Larmor frequency induced by the large static field,
$F_{\mathrm{R}}=55.96\,\mathrm{MHz}$ denotes the Rabi frequency for an isolated spin in a microwave field of $1\,\mathrm{mT}$,
we introduce $h_p$ as a parameter to control the amplitude of the microwave pulse,
$D_0=51.88\,\mathrm{GHz}$, and we express all distances in \AA.
With this choice of units, it is convenient to express frequencies
in MHz and time in $\mu$s.
The new dimensionless variables $\xi^\alpha_j$ for $\alpha=x,y,z$ and $\zeta_j$ are defined by
$g^\alpha_j=g_e(1+\xi^\alpha_j)$ and $\mu_B g_e B'_j/\hbar= 2\pi h_p F_R(1+\zeta_j)$, respectively.
For concreteness, we assume that the spins are located on the Si diamond lattice
with lattice parameter 5.43\AA (to a first approximation, the choice of the lattice is not expected to affect the results).
Not every lattice site is occupied by a spin: We denote the concentration
of spins (number of spins/\AA$^3$) by $\concentration$. In experiment $\concentration\approx10^{-4},\ldots,10^{-6}$.

Guided by experimental results, we assume that the distribution of $\xi^\alpha_j$
is Lorenztian and independent of $\alpha$, cut-off at $\xi_0$, and has a width $\Gamma$:
\begin{eqnarray}
p(\xi^\alpha_j)=\frac{1}{\arctan(\xi_0/\Gamma)}\frac{\Gamma}{(\xi^\alpha_j)^2+\Gamma^2}\Theta(\xi_0-|\xi^\alpha_j|)
.
\label{B5}
\end{eqnarray}
The reason for introducing the cut-off $\xi_0>0$ is that because the Lorentzian distribution has a very long tail,
in practice, we may generate $\xi's$ such that the corresponding value of $g$ is negative, which may not be physical.
Therefore, we use
\begin{eqnarray}
\xi^\alpha_j=\Gamma\tan [(2r-1)\arctan(\xi_0/\Gamma)]
,
\label{B5a}
\end{eqnarray}
to generate the random variables $\xi^\alpha_j$ with distribution Eq.~(\ref{B5})
from uniformly distributed random numbers $0<r<1$.
Likewise, the microwave amplitudes are given by $B'_j=h_p(1+\zeta_j)$ where the $\zeta$'s are random numbers with distribution
\begin{eqnarray}
p(\zeta_j)=\frac{1}{\arctan(\zeta_0/\gamma)}\frac{\gamma}{\zeta^2_j+\gamma^2}\Theta(\zeta_0-|\zeta_j|)
,
\label{B6}
\end{eqnarray}
and $h_p$ is the average amplitude of the microwave field.
Appendix~\ref{parameters} gives a summary of the model parameters that we use in our simulations.

\subsection{Simulation procedure}

The physical properties of interest, in particular the decay rate $c_\mathrm{R}=1/T_\mathrm{R}$ of the Rabi oscillations and
the intrinsic decay rate $c_2=1/T_2$, can be extracted from the time-dependence of
the longitudinal and transverse magnetization, respectively, and are defined by
\begin{eqnarray}
\langle M^z(t) \rangle&=&\sum_{j=1}^L \langle S_j^z \rangle=\sum_{j=1}^L \langle \Psi(t) |S_j^z |\Psi(t)\rangle
\label{B7}
\\
\langle M^x(t) \rangle&=&\sum_{j=1}^L \langle S_j^x \rangle=\sum_{j=1}^L \langle \Psi(t) |S_j^x |\Psi(t)\rangle
,
\label{B8}
\end{eqnarray}
respectively.
We compute the time-dependent wave function $|\Psi(t)\rangle$ by solving the TDSE
\begin{eqnarray}
i\hbar\frac{\partial}{\partial t}|\Psi(t)\rangle=H_{RF}|\Psi(t)\rangle
,
\label{TDSE}
\end{eqnarray}
with $H_{RF}$ given by Eq.~(\ref{M0}).
Numerically, we solve the TDSE using an unconditionally stable product-formula algorithm~\cite{RAED06}.
For the largest spin systems, we perform the simulations using a parallel implementation
of this algorithm, based on a massively parallel quantum computer simulator~\cite{RAED07x}.
Our numerical method strictly conserves the norm of the wave function
and conserves the energy to any desired precision (limited by the machine precision).

In analogy with the experimental procedure, we carry out two types of simulations
yielding the longitudinal (transverse) magnetization $\langle M^z(t)\rangle$
($\langle M^x(t)\rangle$).
From Eq.~(\ref{M0}), it follows directly that
$d \langle M^z(t)\rangle/dt=-\langle M^y(t)\rangle$,
hence $\langle M^z(t)\rangle$ is directly related to $\langle M^y(t)\rangle$ measured in experiment.
We prepare the spin system, that is the state
$|\Psi(0)\rangle$, such that all spins are aligned along the $z$ ($x$)-axis.
Then, for a fixed value of the microwave amplitude $h_p$ ($h_p=0$) and a particular realization
of the random variables $\xi^x_j$, $\xi^y_j$, $\xi^z_j$, $\zeta_j$ and the distribution
of the spins on the lattice, we solve the TDSE and compute Eqs.~(\ref{B7})--(\ref{B8}).
This procedure is then repeated several times with different realizations of random variables
and distributions of spins. Finally we compute the average of Eqs.~(\ref{B7})--(\ref{B8}) over all
these realizations and analyze its time dependence by fitting a simple, damped sinusoidal
function to the simulation data. This then yields the decay rate $c_\mathrm{R}=1/T_\mathrm{R}$
(intrinsic decoherence rate $c_2=1/T_2$) of the Rabi oscillations.

\begin{figure}[t]
\includegraphics[width=8cm]{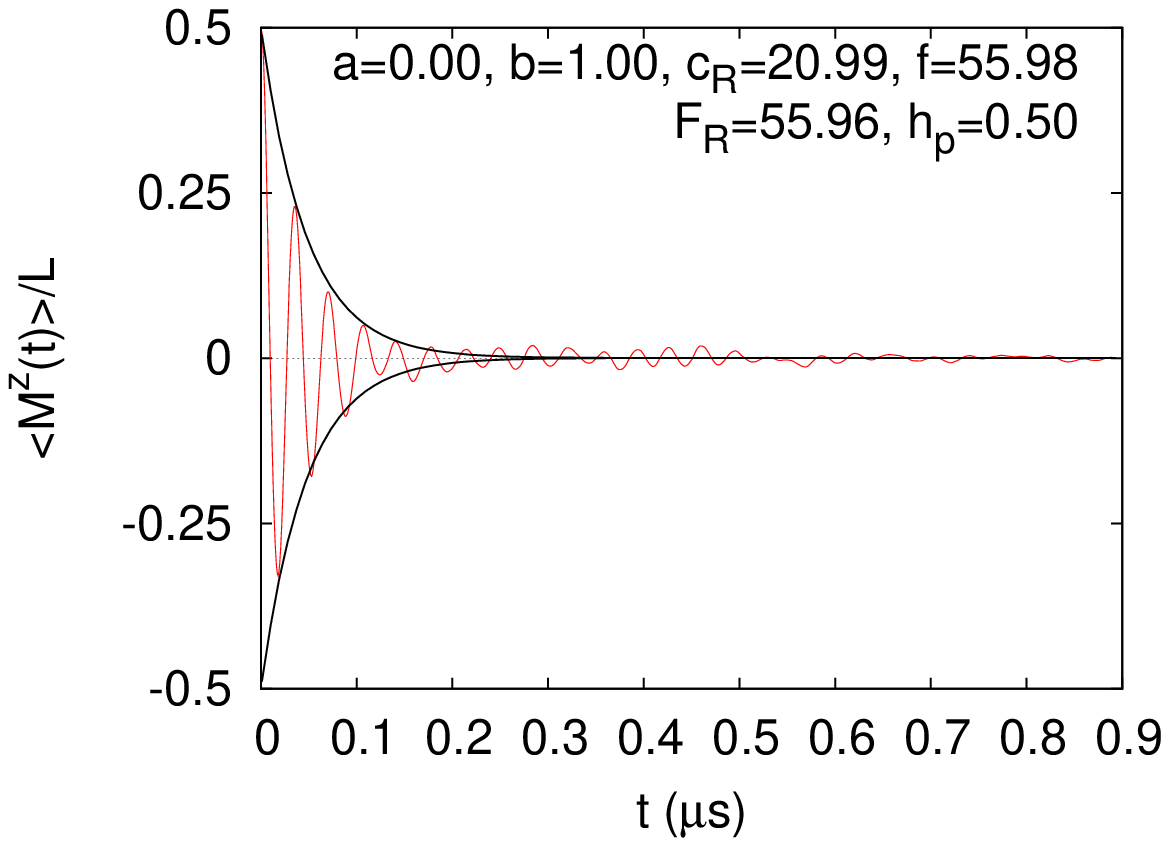} 
\includegraphics[width=8cm]{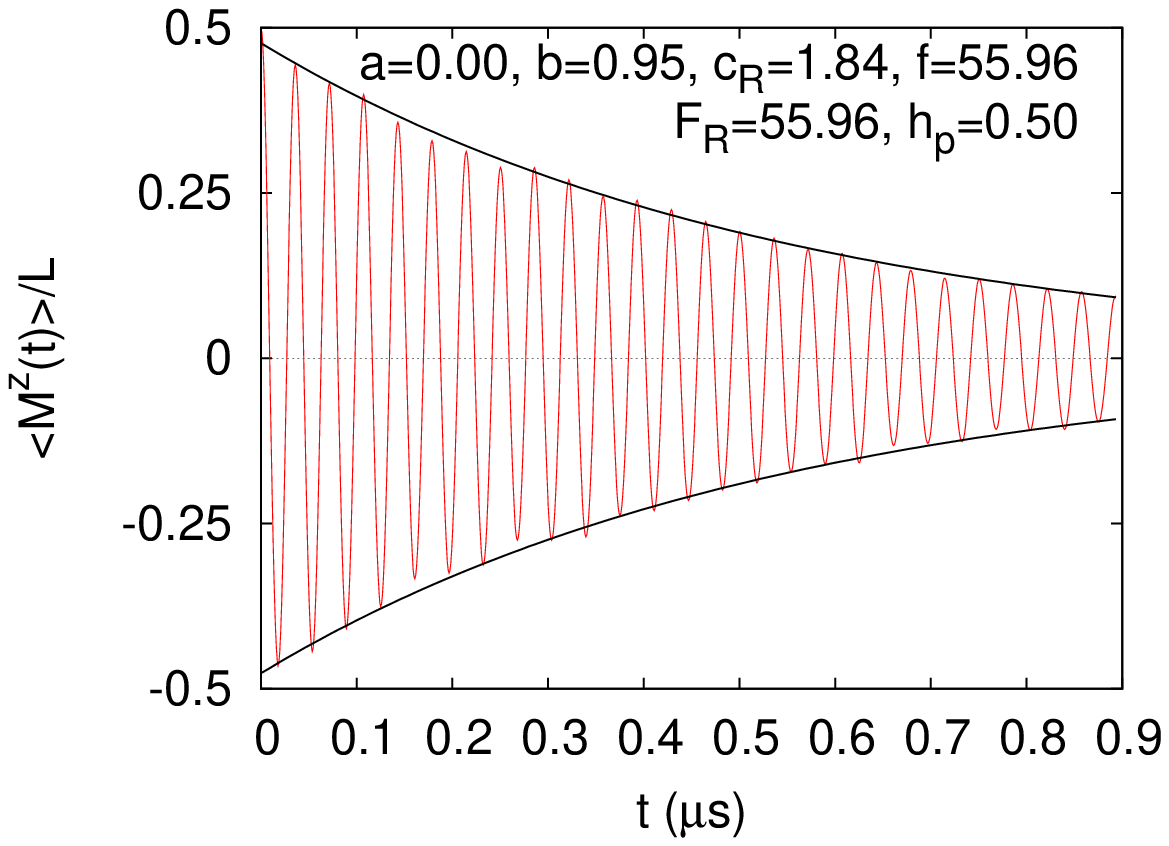}
\includegraphics[width=8cm]{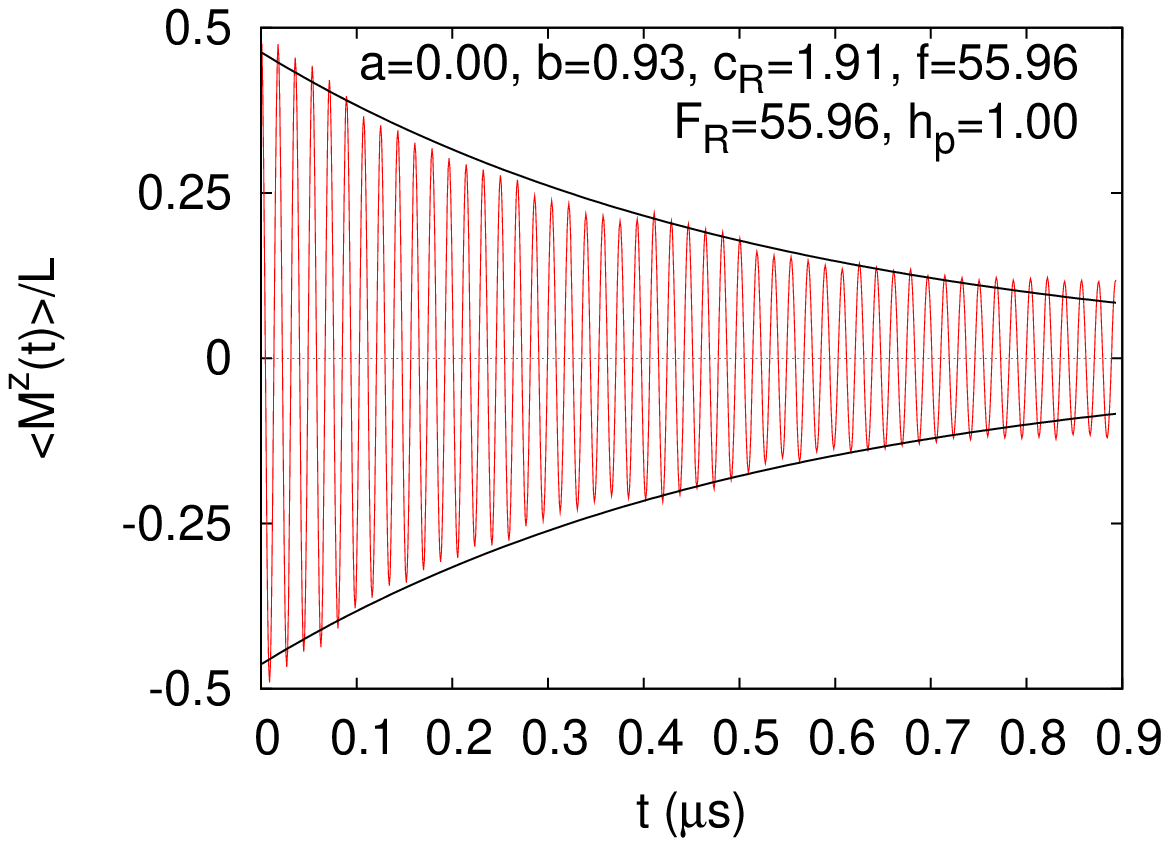}   
\includegraphics[width=8cm]{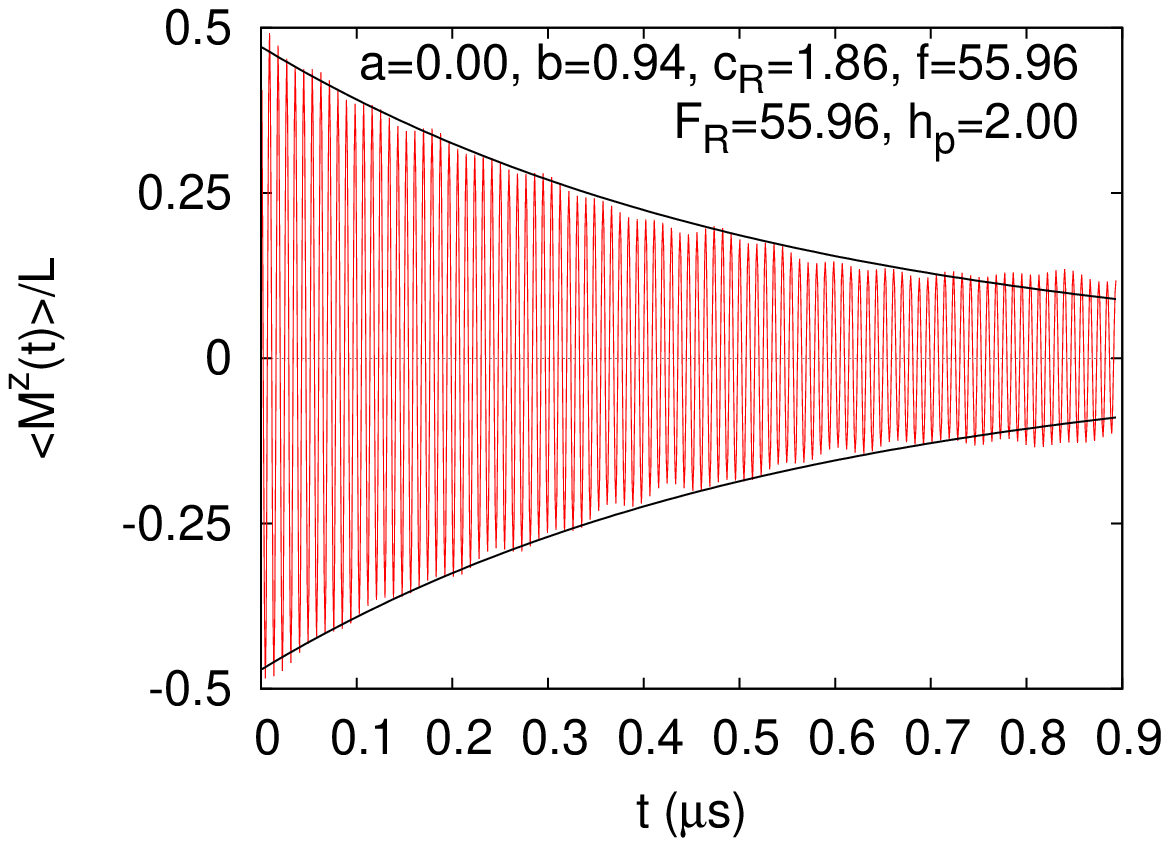}
\caption{(Color online) The Rabi oscillations of the longitudinal magnetization as obtained by solving the TDSE
for the Hamiltonian Eq.~(\ref{M0}) for 26 dipolar-coupled spins
for two different concentrations $\concentration$,
without random fluctuations in the $g$-factors ($\Gamma=0$) and on the microwave amplitude ($\gamma=0$).
Top left: $\concentration=10^{-3}$;
Top right and bottom: $\concentration=10^{-4}$.
The solid lines represents the envelope
$(a\pm be^{-c_\mathrm{R}t})/2$ of the function $(a+be^{-c_\mathrm{R}t}\cos2\pi ft)/2$ that was fitted
to the data.
}
\label{fig.Ab1}
\end{figure}

\begin{figure}[t]
\includegraphics[width=8cm]{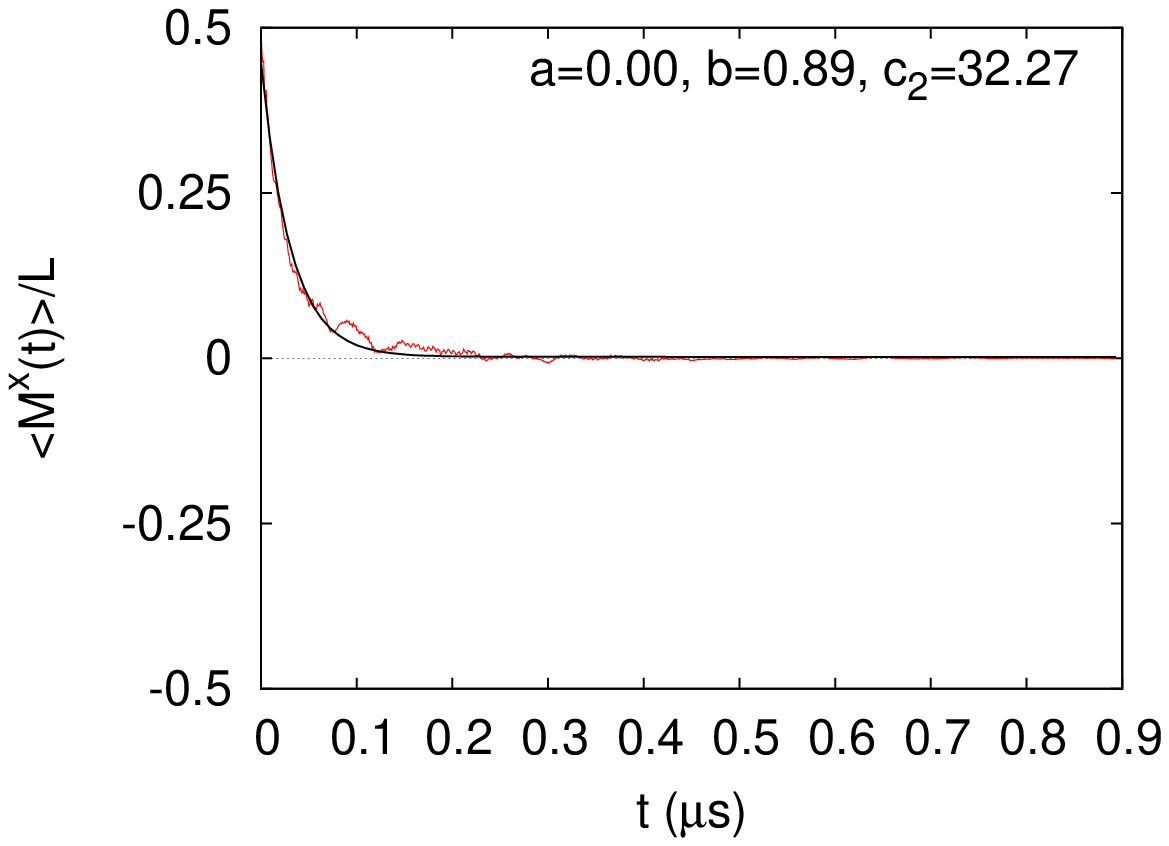}     
\includegraphics[width=8cm]{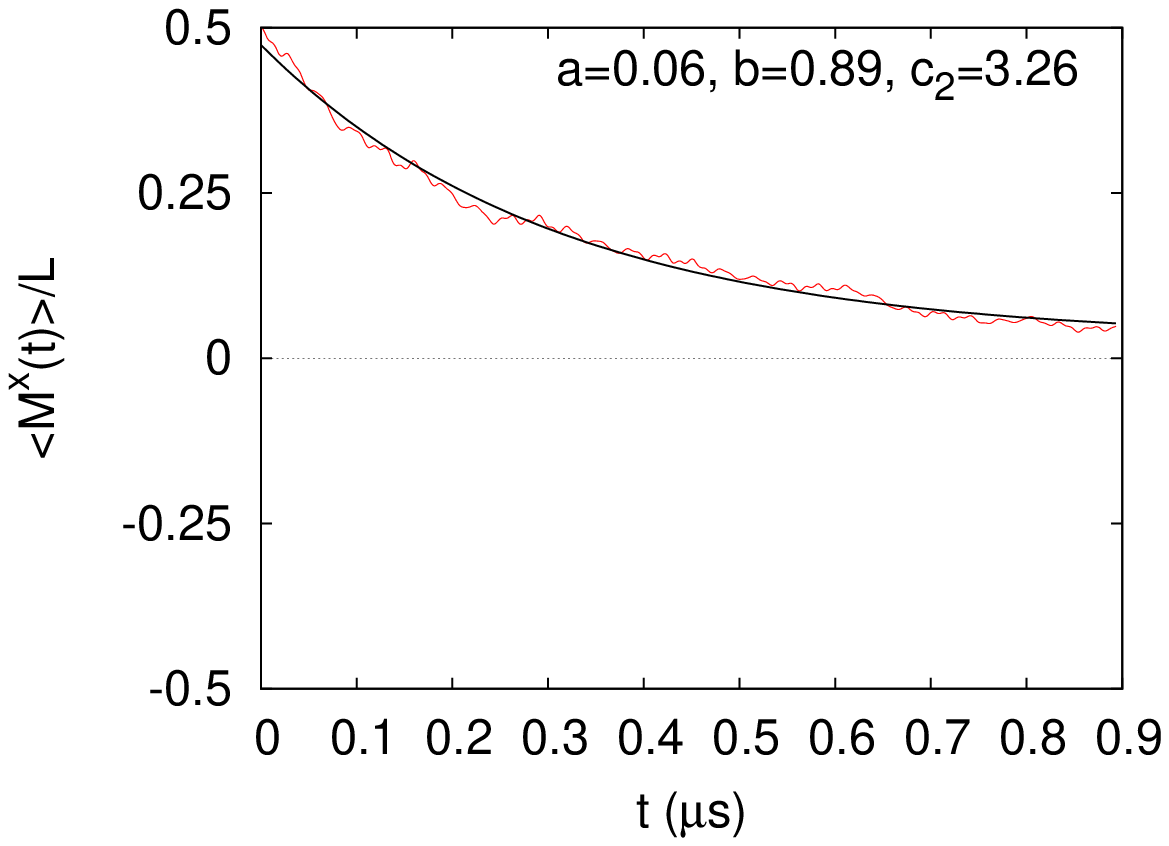}
\caption{(Color online) Time evolution of the transverse magnetization as obtained by solving the TDSE
for the Hamiltonian Eq.~(\ref{M0}) for 26 dipolar-coupled spins
without random fluctuations in the $g$-factors ($\Gamma=0$) and on the microwave amplitude ($\gamma=0$).
Left: $\concentration=10^{-3}$;
Right: $\concentration=10^{-4}$. The non-zero value of $a$ is due to the statistical noise
and the relatively short time interval used for the fit.
The solid line represents the function $(a+ be^{-c_2t})/2$ that was fitted to the data.
}
\label{fig.Ab2}
\end{figure}

\section{Results}\label{results}

In the subsections that follow, we consider the various sources of decoherence separately.
We also study the interplay of intrinsic decoherence due to e.g. pairwise interactions
and extrinsic decoherence due to e.g. single spins driven by external magnetic fields when
different spins have different environments (different couplings to static and microwave fields).
The averaging over different spins leads to decoherence,
that is phase destruction of the electromagnetic waves generated by the spins.
These two types of decoherence lead to the observed damping of Rabi
oscillations which takes place through energy exchange between the spin system and the applied microwave field.
Below we show that energy dissipation from the spin-bath to the electromagnetic bath is sufficient
to explain the experimental results on the Rabi decay time.
This is the reason why we neglect, in the present paper,
the dissipation effect of phonons, (our spin-lattice relaxation time $T_1$ is infinite).
Note that if we turn off the microwave field,
the longitudinal component of the magnetization commutes with the Hamiltonian Eq.~(\ref{M0})
and hence does not change with time at all,
showing that energy exchange with the electromagnetic bath is essential.
In the following sections we give two examples in which energy flows from
the electromagnetic bath to the spin-bath and
from the spin-bath to the electromagnetic bath.

\subsection{Fixed $g$-factors and homogeneous fields}\label{A}
In the absence of randomness on the $g$-factors or on the microwave amplitude, the Hamiltonian is given by Eq.~(\ref{M0})
with $\xi_x=\xi_y=\xi_z=\zeta=0$.

\subsubsection{Non-interacting spins}\label{Aa}
For non-interacting spins, we can drop the spin label and write the Hamiltonian (in the rotating frame)
for a single spin as
\begin{eqnarray}
H_{RF}/\hbar&=&- 2\pi h_p F_{\mathrm{R}} S^x
.
\label{Eq.Aa.1}
\end{eqnarray}
The time evolution of the longitudinal magnetization takes the simple form
\begin{eqnarray}
\langle\Psi(t)|S^z|\Psi(t)\rangle&=&\frac{1}{2}\cos \Omega_{\mathrm{R}}t
,
\label{Eq.Aa.2}
\end{eqnarray}
showing that the $z$-component of the spin performs undamped Rabi oscillations
with angular frequency $\Omega_{\mathrm{R}}=2\pi h_p F_{\mathrm{R}}$.
Therefore $T_{\mathrm{R}}=\infty$.
Furthermore, the transverse magnetization is conserved and therefore $T_2=\infty$.
Summarizing, in the absence of randomness and dipole-dipole interactions, we have
\begin{equation}
T_{\mathrm{R}}=\infty
\quad,\quad
T_2=\infty
.
\label{Eq.Aa.3}
\end{equation}

\begin{figure}[t]
\includegraphics[width=14cm]{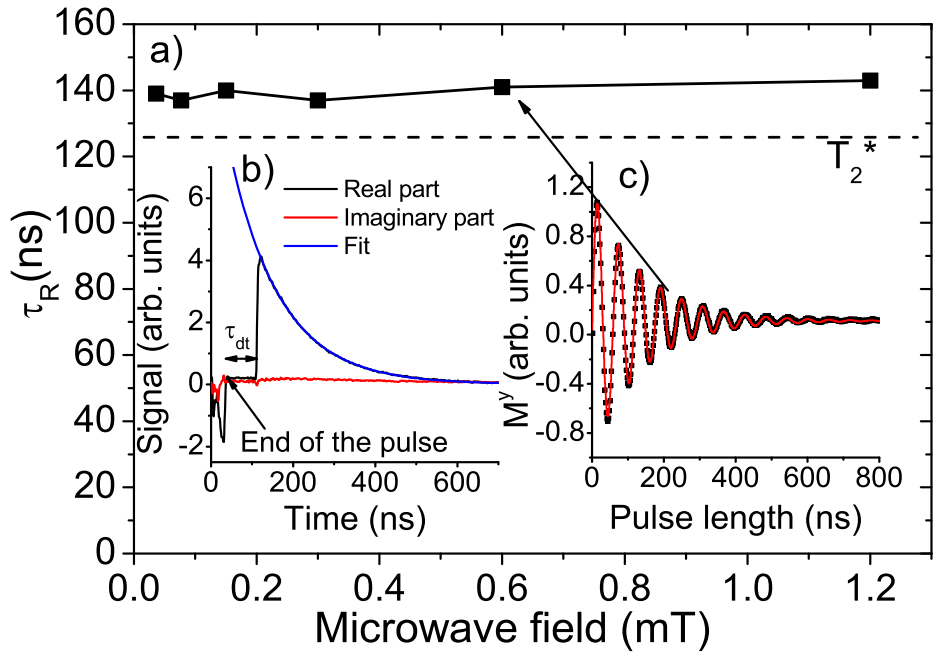}
\caption{(Color online)
a) Decay time $\tauR $ of the Rabi oscillation of BDPA as a function of
the amplitude of microwave field. Dashed line: Value of $T_2^\ast$ the decay time of the FID.
b) Example of the FID signal. $\tau_{dt}$: Dead time of the spectrometer.
Black line (red line): In-phase (out-of-phase)
FID signal recorded by the detector.
Blue line: Best fit to the exponential decay.
c) Example of Rabi oscillations. The red line is the best
fit to $M^y(t) = A_0 \sin(\Omega_R t+\phi)\exp(-t/\tauR )+M^y(\infty)$,
giving the Rabi decay time $\tauR $ and a non-zero offset.
The nonzero value of $M^y(\infty)$ is due to dissipation effects~\cite{weil1994}, collectively described by the relaxation time $T_1$,
which are not included in the microscopic model considered in the present paper.
Measurements were carried out at room temperature.
}
\label{fig:bdpa} \end{figure}

\subsubsection{Dipolar-coupled spins}\label{Ab}

In Figs.~\ref{fig.Ab1} and \ref{fig.Ab2}, we present simulation results
for the longitudinal and transverse magnetization, respectively, as obtained by averaging
the solutions of the TDSE over ten different distributions of 26 dipolar-coupled spins on the lattice.
Our simulation results, many of them not shown, lead us to the following conclusions:
\begin{itemize}
\item{For both concentrations $\concentration=10^{-3}$ and $\concentration=10^{-4}$
and for microwave amplitudes $h_p=0.5,1,2$, the Rabi oscillations decay exponentially.
Indeed, the fits are good, as indicated by the small differences between the Rabi frequency ($F_0=55.96$)
and the values of $f$ obtained by the fitting procedure.
}
\item{The decay rate $c_\mathrm{R}=1/T_\mathrm{R}$ increases with $\concentration$, with a slope of
approximately 1.7 (data not shown).
}
\item{Within the statistical fluctuations resulting from the random distribution of the spins
on the lattice, $c_\mathrm{R}=1/T_\mathrm{R}$ does not depend on the microwave amplitude $h_p$
but strongly depends on the concentration $n$.
}
\item{
Simulations (data not shown) for $\concentration=0.25\times10^{-4},\ldots,10^{-3}$
indicate that $T_\mathrm{2}\propto \concentration$, as expected theoretically.
}
\item{The simulation data suggest that $c_2=1/T_2 > c_\mathrm{R}=1/T_\mathrm{R}$.
}
\end{itemize}
Summarizing, in the absence of local randomness but in the presence of dipole-dipole interactions,
we have
\begin{equation}
T_{\mathrm{R}}=T_{\mathrm{R}}(\concentration)>T_{\mathrm{2}}=T_{\mathrm{2}}(\concentration)
.
\label{Eq.Ab.1}
\end{equation}

\subsubsection{Experimental results: BDPA}

We now compare these theoretical predictions to experiments performed
on a single crystal of BDPA ($\alpha-\gamma$-bisdiphenylene-$\beta$-phenylally).
With a linewidth of $0.09\,$mT, this
system is quite homogeneous with a very narrow distribution of the $g$-factors.
Moreover, the sample used was very tiny such that we may consider the microwave to be homogeneous
inside the sample.

Results are presented in Fig.~\ref{fig:bdpa}.
They show an example of Rabi oscillations obtained from FID measurements.
The Rabi oscillations fit very well to
\begin{eqnarray}
 M^y(t) &=&
A_0 \sin(\Omega_R t+\phi)\exp(-t/\tauR )+M^y(\infty),
\label{eq:RabiX}
\end{eqnarray}
for all microwave powers.
The obtained Rabi decay time $\tauR $
is clearly independent of the amplitude of the microwave field, as predicted by the model when $\Gamma=\gamma=0$.
It is also very close to $T_2^\ast$, the FID decay time given by the Fourier transform of the EPR linewidth.
This is also in agreement with predictions when  $\Gamma=\gamma=0$ and $D_0\not=0$,
$T_2^\ast$ being a coherence time fully equivalent to $T_2$.
The discrepancy between $\tauR $ ($\sim 140$~ns)
and $T_2^*$ ($= 128$~ns) is due to a small inhomogeneous broadening (about 10\%).

\begin{figure}[t]
\includegraphics[width=8cm]{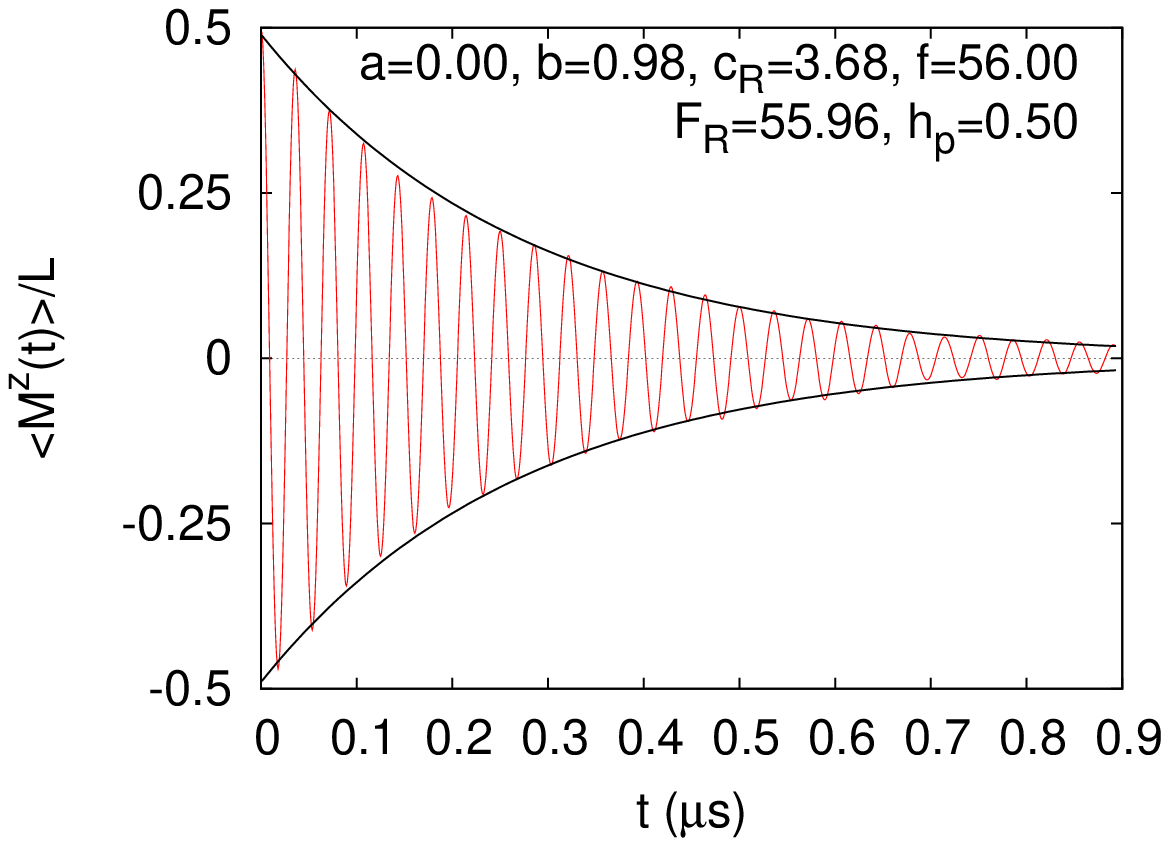}
\includegraphics[width=8cm]{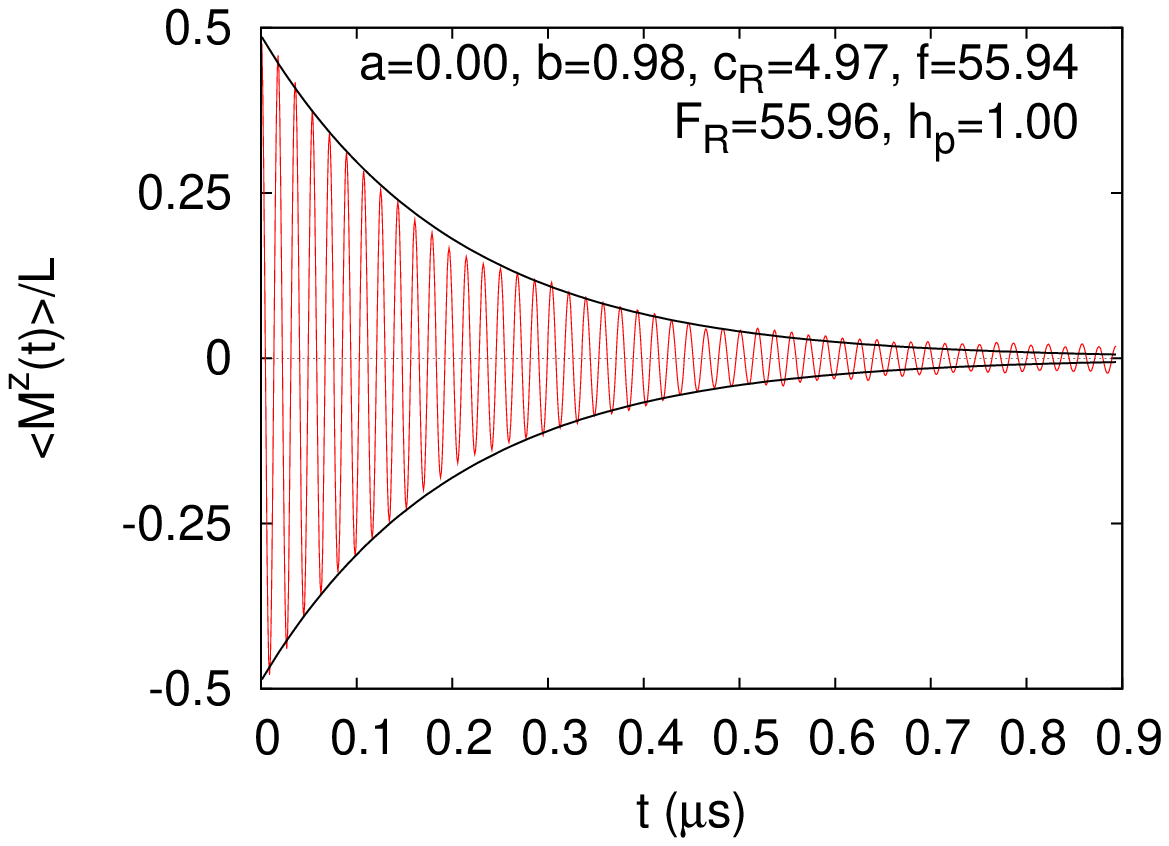}
\includegraphics[width=8cm]{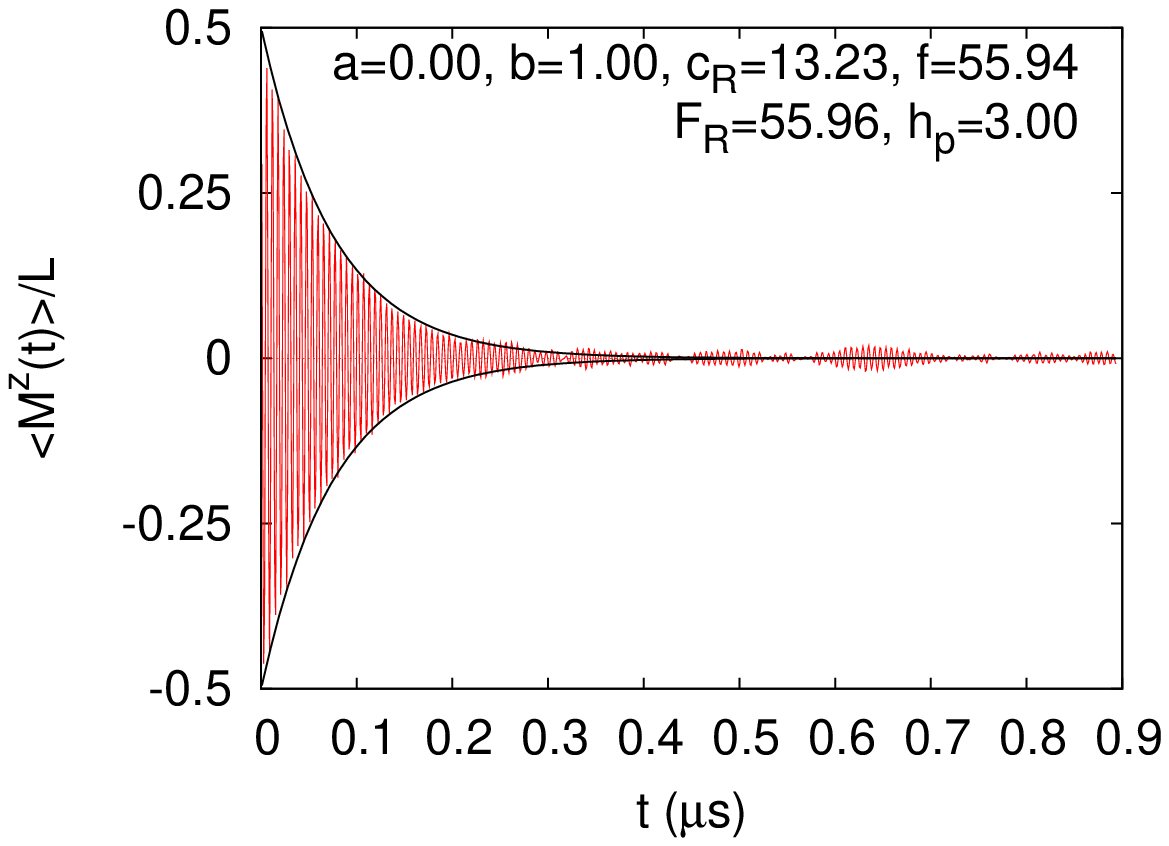}
\includegraphics[width=8cm]{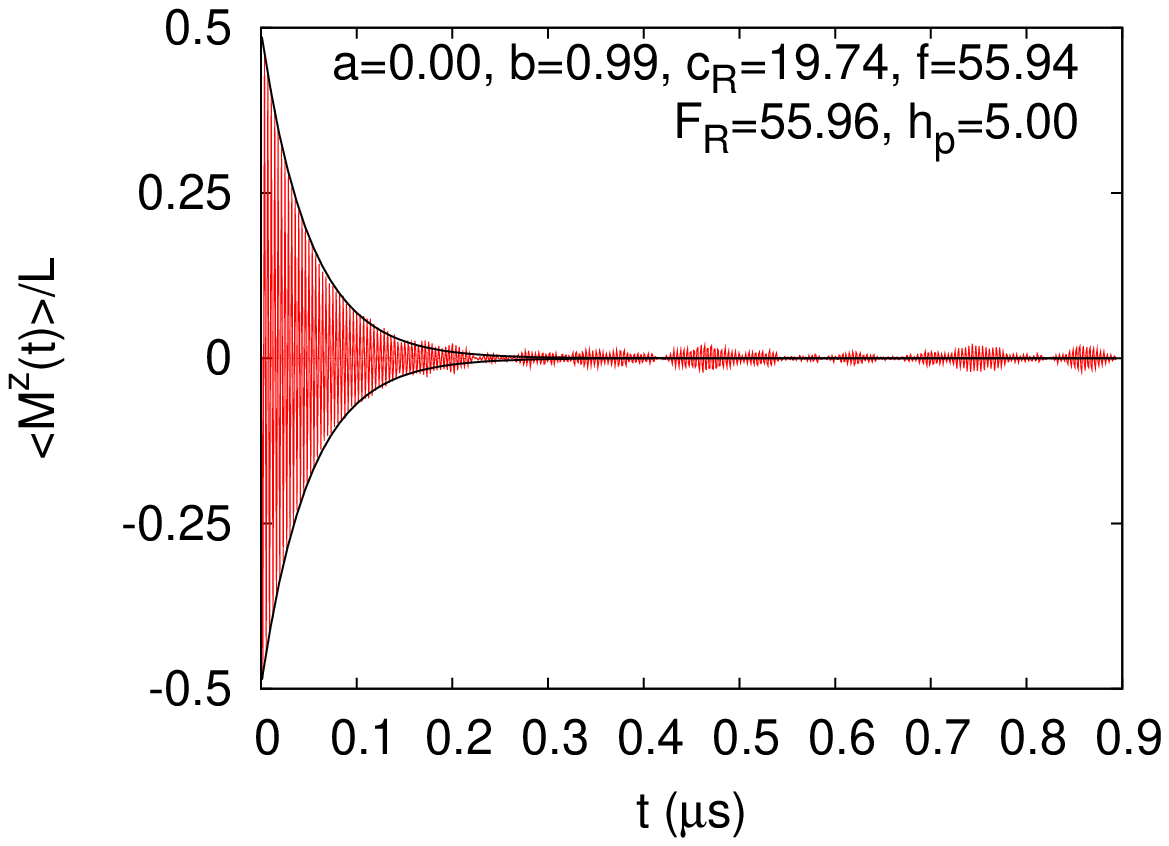}
\includegraphics[width=8cm]{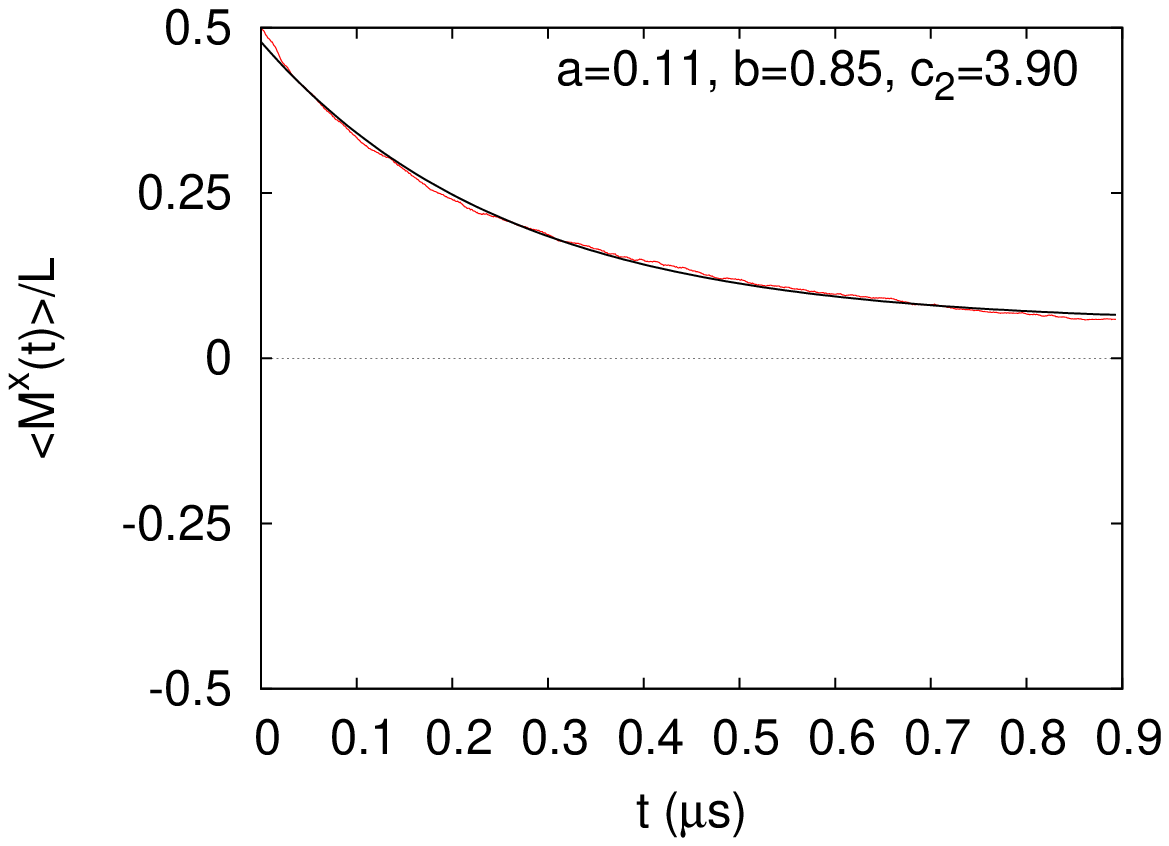}
\includegraphics[width=8cm]{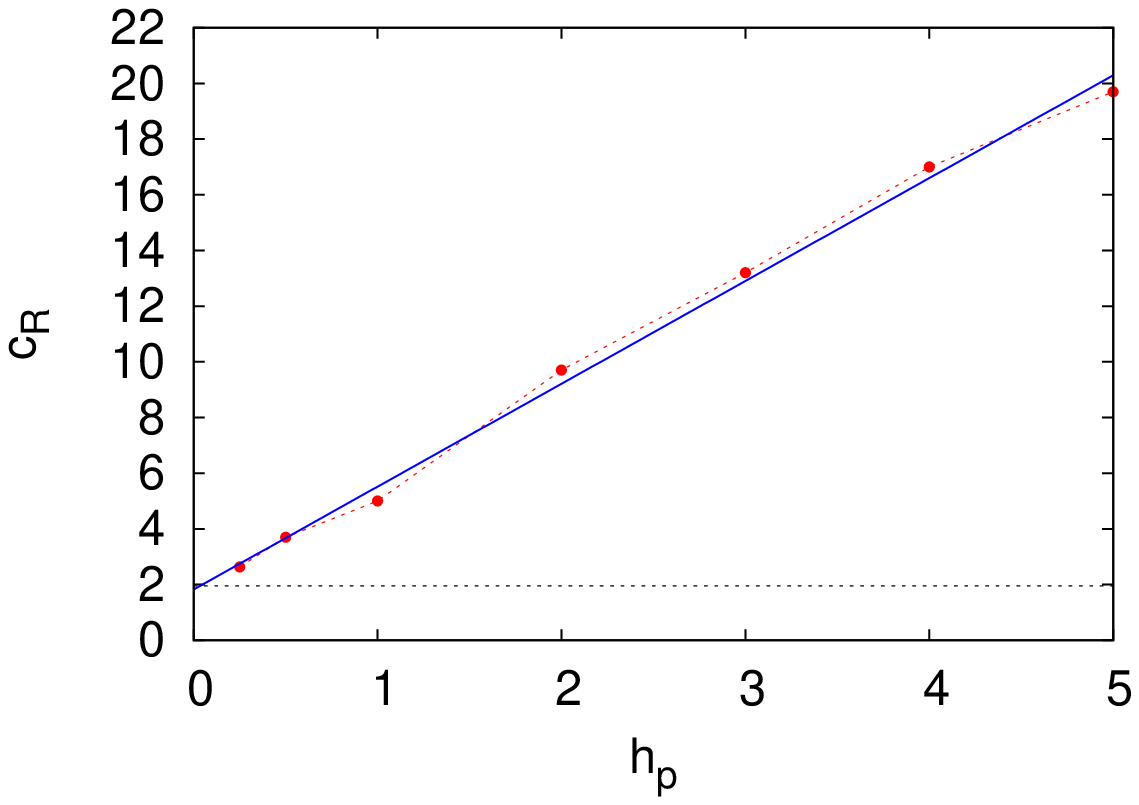}
\caption{(Color online) Simulation results as obtained by solving the TDSE
for the Hamiltonian Eq.~(\ref{M0}) for 12 spins (concentration $\concentration=10^{-4}$)
that interact via dipole-dipole interaction,
without random fluctuations in the $g$-factors ($\Gamma=0$) but with random fluctuations on the microwave amplitude ($\gamma=0.01$).
The results represent the average of 100 different realizations of 12-spin systems.
Top left to middle right: Longitudinal magnetization for different values of $h_p$.
The solid line represents the envelope $(a\pm be^{-c_\mathrm{R}t})/2$ of the function $(a+be^{-c_\mathrm{R}t}\cos2\pi ft)/2$ that was fitted to the data.
Bottom left: Transverse magnetization in the absence of the microwave field ($h_p=0$).
The solid line represents the function $(a+ be^{-c_2t})/2$ that was fitted to the data.
Bottom right:
Bullets show the inverse relaxation time $c_\mathrm{R}=1/T_\mathrm{R}$ as a function of the microwave amplitude $h_p$.
The dashed line connecting the bullets is a guide to the eye only.
A linear fit to the simulation data yields $c_\mathrm{R}=1/T_\mathrm{R}\approx 3.69 h_p + 1.82$ and is shown by the solid line.
The horizontal line represents the value of $c_2=1/2T_2\approx 1.95$, estimated from the data of the transverse magnetization (see bottom left).
}
\label{fig.Bb}
\end{figure}

\begin{figure}[t]
\includegraphics[width=12cm]{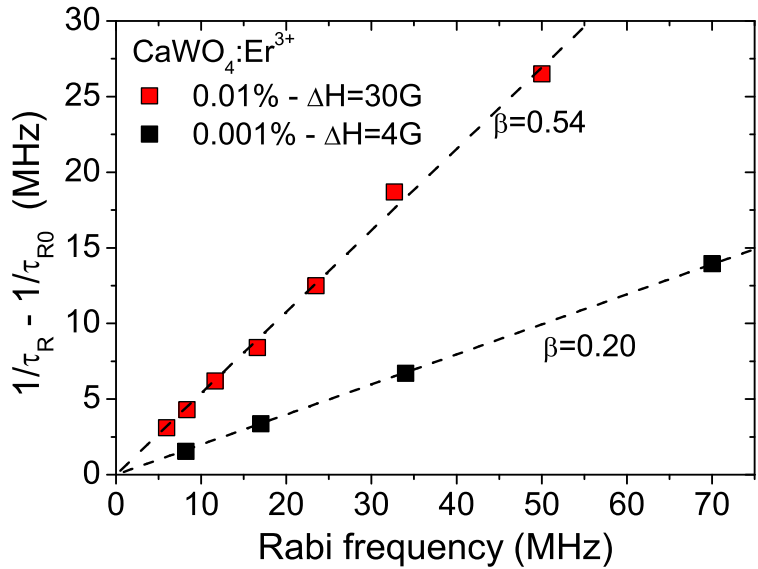}
\caption{(Color online)
Decay time $\tauR $ of Rabi oscillations in CaWO$_4$
as a function of the microwave field amplitude (corresponding to the Rabi frequency $\Omega_R$)
for two concentrations of the Er$^{3+}$ spins.
The static field $H_0$ is parallel to the $c$-axes of the cristal and the temperature T$\sim$4~K.
Dashed lines: Fit to  $1/\tauR=1/\tauRR  +\beta \Omega_R/2\pi$
Measurements were carried out at $4\,\mathrm{K}$.
}
\label{fig:Er}
\end{figure}

\begin{figure}[t]
\includegraphics[width=12cm]{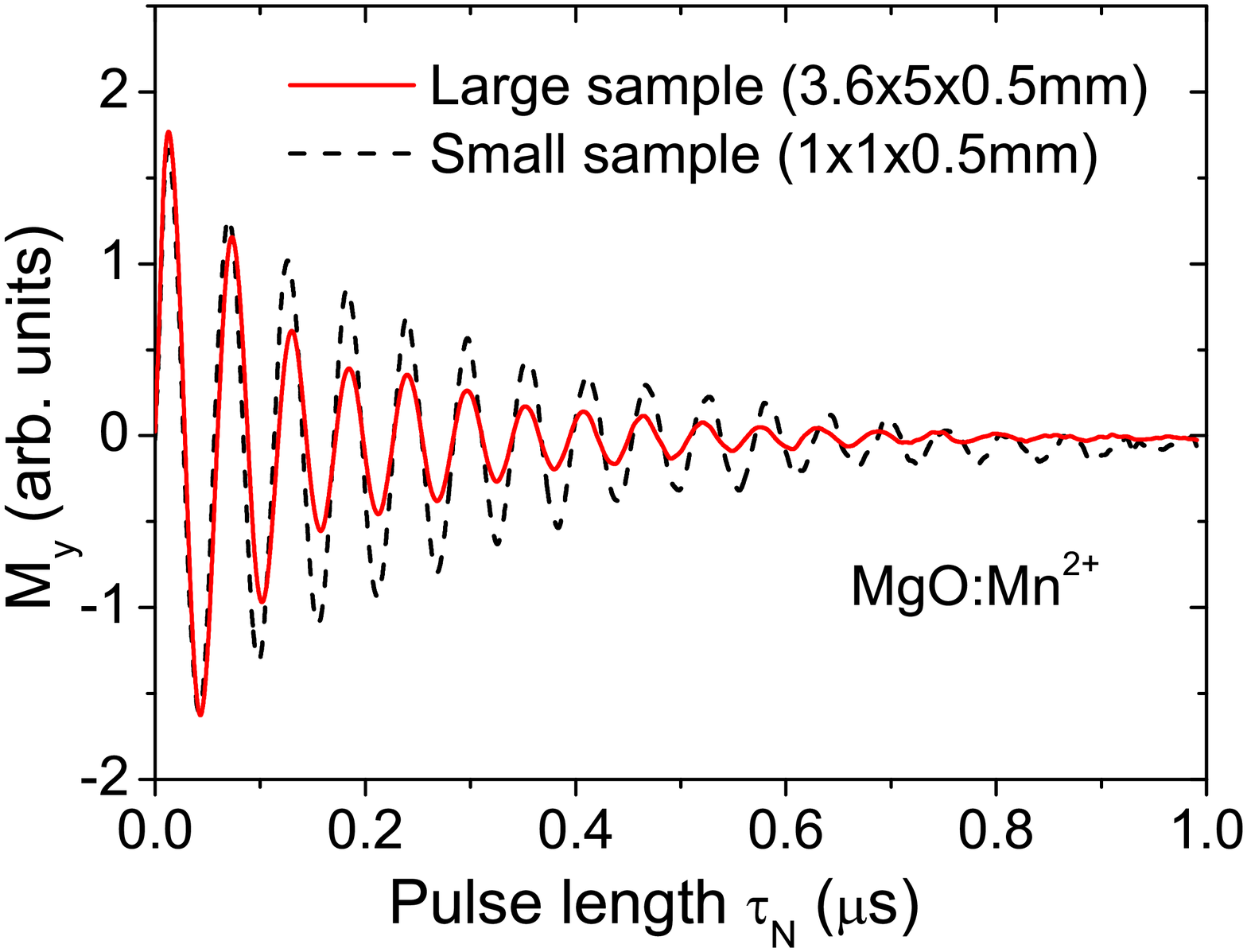}
\caption{(Color online)
Rabi oscillations of MgO:Mn$^{2+}$ (0.001\%) for two sample sizes.
Measurements were carried out at room temperature.
}
\label{fig:Mn1}
\end{figure}

\subsection{Randomness in the microwave amplitude only}\label{B}
In the case of randomness in the microwave amplitude only, the Hamiltonian is given by Eq.~(\ref{M0}) with $\xi^x=\xi^y=\xi^z=0$.
Such a randomness is inherent to finite size cavities and becomes smaller as the size of the sample relative
to the size of the cavity is reduced.

\subsubsection{Non-interacting spins}\label{Ba}
For non-interacting spins ($D_0=0$), we
can readily compute the average over the distribution of $\zeta_j$ analytically if
we neglect the cut-off of the Lorentzian distribution.
As all spins are equivalent, we may drop the spin index $j$
and we obtain
\begin{eqnarray}
\overline{\langle S^z(t)\rangle}
&=&
\frac{\gamma}{2\pi} \int_{-\infty}^{+\infty}
\frac{\cos \Omega_{\mathrm{R}}(1+\zeta)t}{\zeta^2+\gamma^2} \,d\zeta
=\frac{1}{2}e^{-\gamma \Omega_{\mathrm{R}} t}\cos \Omega_{\mathrm{R}} t
,
\label{Eq.Ba.1}
\end{eqnarray}
showing that the Rabi oscillations decay exponentially
and that the decay time of the Rabi oscillations is given by
$1/T_{\mathrm{R}}=\gamma \Omega_{\mathrm{R}}$.
Furthermore, the transverse magnetization is conserved and therefore $T_2=\infty$.
Summarizing, in the presence of randomness
in the microwave field only and in the absence of dipole-dipole interactions, we have
\begin{equation}
1/T_{\mathrm{R}}=\gamma \Omega_{\mathrm{R}} > 1/T_{\mathrm{2}}=0
,
\label{Eq.Ba.2}
\end{equation}
showing that the decay rate of the Rabi oscillations increases linearly
with the microwave amplitude $h_p$ whereas $T_2$ remains infinite.
This is easy to understand: $T_2$ is infinite due to the lack of
pairwise intrinsic decoherence whereas destructive interference associated
with weak positional randomness in $h_p$ (the microwave field)
leads to a reduction of $T_R$ when $h_p$ increases (one-qubit decoherence).

\subsubsection{Interacting spins: dipole-dipole interaction}\label{Bb}

In Fig.~\ref{fig.Bb}, we present simulation results for systems of 12 spins with dipole-dipole interaction
and randomness in $h_p$,
as obtained by averaging over 100 different realizations, meaning 100 different distributions of the 12 spins on the lattice.
The four upper panels of Fig.~\ref{fig.Bb} show results for the longitudinal magnetization
$\langle M^z(t) \rangle$.

Rabi oscillations are damped but have zero offset.
The inverse Rabi time $1/\tauR =c_R$,
deduced from sinusoidal fits, increases linearly with the microwave field, that is with the Rabi frequency $\Omega_R$
(bottom right).
Its value at $h_p=0$ is to good accuracy equal to $1/2T_2$ ($a_0\approx1.82$ for $n=10^{-4}$).
The slope $a_1\approx3.69/F_R$
is related to the matrix of the gyromagnetic factor
and to the root mean square of local fields resulting from the randomness in the microwave field.
These results, specific to a $h_p$ distribution,
agree qualitatively with recently published results
of the damping time of Rabi oscillations in the limit of a large inhomogenous linewith~\cite{Baibekov11a}.

The results for the transverse magnetization $\langle M^x(t) \rangle$
in the absence of microwaves ($h_p=0$)
are presented in the bottom left panel of Fig.~\ref{fig.Bb}.
It clearly decays exponentially, as this is the case with the longitudinal magnetization.
Summarizing, from Fig.~\ref{fig.Bb} we conclude that in the presence of randomness
in the microwave field and of dipole-dipole interactions, we have
\begin{equation}
c_\mathrm{R}=1/T_\mathrm{R}\approx a_1 \Omega_\mathrm{R}  + a_0
\quad,\quad
a_0\approx1/2T_2
.
\label{Eq.Bb.1}
\end{equation}
Here,  pairwise decoherence affects $T_2$
which is now finite (and proportional to $1/n$ as in the case without randomness, see Section~\ref{A})
and randomness in microwave amplitude $h_p$ affects $T_R$ which is essentially proportional
to $1/h_p$ at large $h_p$.
As $T_R<T_2$, we can say that, in this case,
energy flows from the spin-bath to the electromagnetic bath,
leading to energy dissipation in the spin-bath.

\subsubsection{Experimental results: CaWO$_4$:Er$^{3+}$ and MgO:Mn$^{2+}$.}

In order to show the effect of concentration on Rabi damping we measure two samples of CaWO$_4$:Er$^{3+}$
with Erbium concentration 0.01\% and 0.001\%, respectively.
The two samples have nearly the same shape, keeping the inhomogeneity of microwave field constant.
To remove the effects of zero microwave field decay (that is $T_2$ due to multi-spin or pairwise decoherence)
we plot $1/\tauR -1/\tauRR  $ where $\tauRR  \approx 1/2T_2$ is the decay time at zero microwave field.
The results are presented in Fig.~\ref{fig:Er}.
The inverse Rabi decay time fits very well to $1/\tauR=1/\tauRR  +\beta \Omega_R/2\pi$,
where $\beta$ is a fitting parameter.
From Fig.~\ref{fig:Er}, it is clear that the Rabi-decay time $\tauR $ decreases
with the concentration $n$, in concert with the simulation results.

Evidence of the effect of microwave field inhomogeneity on the Rabi oscillation decay
has been recently given for a sample of Cr:CaWO$_4$~\cite{Baibekov11b}.
To provide further evidence, we took a sample of MgO doped with about 0.001\% with Mn$^{2+}$
and cut the sample into a large ($3.6\times5\times0.5\,\mathrm{mm}^3$) and small ($1\times1\times0.5\,\mathrm{mm}^3$) piece.
At this extremely low concentration, the dipole-dipole interaction effect on the Rabi decay
is  negligible, hence disorder essentially due to the microwave field inhomogeneity inside these samples will be different.
Fig. \ref{fig:Mn1} shows the Rabi oscillations for these two samples.
All parameters (microwave power, temperature, crystal orientation) are the same
for the measurements on these two samples.
The effect of the inhomogeneity of the microwave field on the Rabi decay time is clearly seen
as the damping in the large sample (red line) is almost two times larger than the one in the small sample (black line).

\subsection{Randomness in the $g$-factors only}\label{C}

We assume that there are no random fluctuations in the amplitude of the microwave pulse
and that the $g$-factors fluctuate randomly from spin to spin.
This effect is generally due to weak crystal distortions, imperfections,
leading to small variations of crystal-field parameters.

\subsubsection{Randomness in $g_z$: Non-interacting spins}\label{Ca1}
In this case, the Hamiltonian is given by Eq.~(\ref{M0}) with $\zeta_j=\xi^x=\xi^y=D_0=0$.
As we then have a system of independent spins, we may drop the spin index $j$.
In the case that initially, all the spins are aligned along the $z$-axis, we find
\begin{eqnarray}
\overline{\langle S^x(t)\rangle}
&=&0
,
\nonumber \\
\overline{\langle S^y(t)\rangle}
&=&-
\frac{1}{2} \int_{-\infty}^{+\infty} 
\frac{\ h_pF_\mathrm{R}
\sin \left[ 2\pi t\sqrt{(F_0\xi^z)^2+(\ h_pF_\mathrm{R})^2}\right]}{\sqrt{(F_0\xi^z)^2+(\ h_pF_\mathrm{R})^2}}
p(\xi^z)
\,d\xi^z
\nonumber \\
&=&
-\frac{(\ h_pF_\mathrm{R}}{2}
\left\{
\frac{\sin \left[2\pi t \sqrt{(\ h_pF_\mathrm{R})^2-(\Gamma F_0)^2}\right]}{\sqrt{(\ h_pF_\mathrm{R})^2-(\Gamma F_0)^2}}
\right.
\nonumber \\
&&-
\left.
2\pi\Gamma F_0
\int_0^t
J_0(2\pi \ h_pF_\mathrm{R} u)
\frac{\sin \left[2\pi(t-u)\sqrt{(\ h_pF_\mathrm{R})^2-(\Gamma F_0)^2}\right]}{\sqrt{(\ h_pF_\mathrm{R})^2-(\Gamma F_0)^2}}
\,d u
\right\}
,
\nonumber \\
\overline{\langle S^z(t)\rangle}
&=&
\frac{1}{2} \int_{-\infty}^{+\infty} 
\frac{(F_0\xi^z)^2+(h_pF_\mathrm{R})^2
\cos \left[ 2\pi t\sqrt{(F_0\xi^z)^2+(h_pF_\mathrm{R})^2}\right]}{(F_0\xi^z)^2+(h_pF_\mathrm{R})^2}
p(\xi^z)
\,d\xi^z
\nonumber \\
&=&
\frac{1}{2}
\left\{
\frac{-(\Gamma F_0)^2+(h_p F_\mathrm{R})^2
\cos \left[2\pi t \sqrt{(h_pF_\mathrm{R})^2-(\Gamma F_0)^2}\right]}{(h_pF_\mathrm{R})^2-(\Gamma F_0)^2}
\right.
\nonumber \\
&&+
\left.
2\pi\Gamma F_0 (h_p F_\mathrm{R})^2
\int_0^t
J_0(2\pi h_pF_\mathrm{R} u)
\frac{1-\cos \left[2\pi(t-u)\sqrt{(h_pF_\mathrm{R})^2-(\Gamma F_0)^2}\right]}{(h_pF_\mathrm{R})^2-(\Gamma F_0)^2}
\,d u
\right\}
.
\label{Eq.Ca1.1}
\end{eqnarray}

In the case that initially, all the spins are aligned along the $x$-axis, we find
\begin{eqnarray}
\overline{\langle S^x(t)\rangle}
&=&
\frac{1}{2} \int_{-\infty}^{+\infty} 
\frac{(h_pF_\mathrm{R})^2+(F_0\xi^z)^2
\cos \left[ 2\pi t\sqrt{(F_0\xi^z)^2+(h_pF_\mathrm{R})^2}\right]}{(F_0\xi^z)^2+(h_pF_\mathrm{R})^2}
p(\xi^z)
\,d\xi^z
\nonumber \\
&=&
\frac{1}{2}
\left\{
\frac{(h_p F_\mathrm{R})^2-(\Gamma F_0)^2
\cos \left[2\pi t \sqrt{(h_pF_\mathrm{R})^2-(\Gamma F_0)^2}\right]}{(h_pF_\mathrm{R})^2-(\Gamma F_0)^2}
-2\pi
\frac{\Gamma F_0  (h_p F_\mathrm{R})^2}{(h_pF_\mathrm{R})^2-(\Gamma F_0)^2}
\int_0^t
J_0(2\pi h_pF_\mathrm{R} u)
\,d u
\right.
\nonumber \\
&&
\left.
+
2\pi
\frac{(\Gamma F_0)^3}{(h_pF_\mathrm{R})^2-(\Gamma F_0)^2}
\int_0^t
J_0(2\pi h_p F_\mathrm{R} u)\cos \left[2\pi (t-u)\sqrt{(h_pF_\mathrm{R})^2-(\Gamma F_0)^2}\right]
\,d u
\right\}
.
\label{Eq.Ca1.2}
\end{eqnarray}
Recall that we calculate the transverse magnetization for the case that initially, all spins are aligned along the $x$-axis.
In order to obtain the expressions in terms of elementary functions, we have ignored the cut-off of the Lorentzian distribution.
We can check that for $\Gamma=0$, Eq.~(\ref{Eq.Ca1.1}) and Eq.~(\ref{Eq.Ca1.2}) reduce to
\begin{eqnarray}
\overline{\langle S^z(t)\rangle}=\frac{1}{2} \cos \Omega_\mathrm{R}t
\quad&,&\quad
\overline{\langle S^x(t)\rangle}=\frac{1}{2}
,
\label{Eq.Ca1.3}
\\
\noalign{\noindent while for $h_p=0$, we find}
\overline{\langle S^z(t)\rangle}=\frac{1}{2}
\quad&,&\quad
\overline{\langle S^x(t)\rangle}
=\frac{1}{2}e^{-2\pi t \Gamma F_0}
,
\label{Eq.Ca1.4}
\end{eqnarray}
in agreement with the expressions that can be derived directly, without any averaging procedure.
From Eq.~(\ref{Eq.Ca1.4}), it follows that $1/T_2= 2\pi \Gamma F_0$.
For finite $\Gamma$, Rabi oscillations are present only if $h_p F_R>\Gamma F_0$
in both longitudinal and transverse cases.
\begin{figure}[t]
\includegraphics[width=8cm]{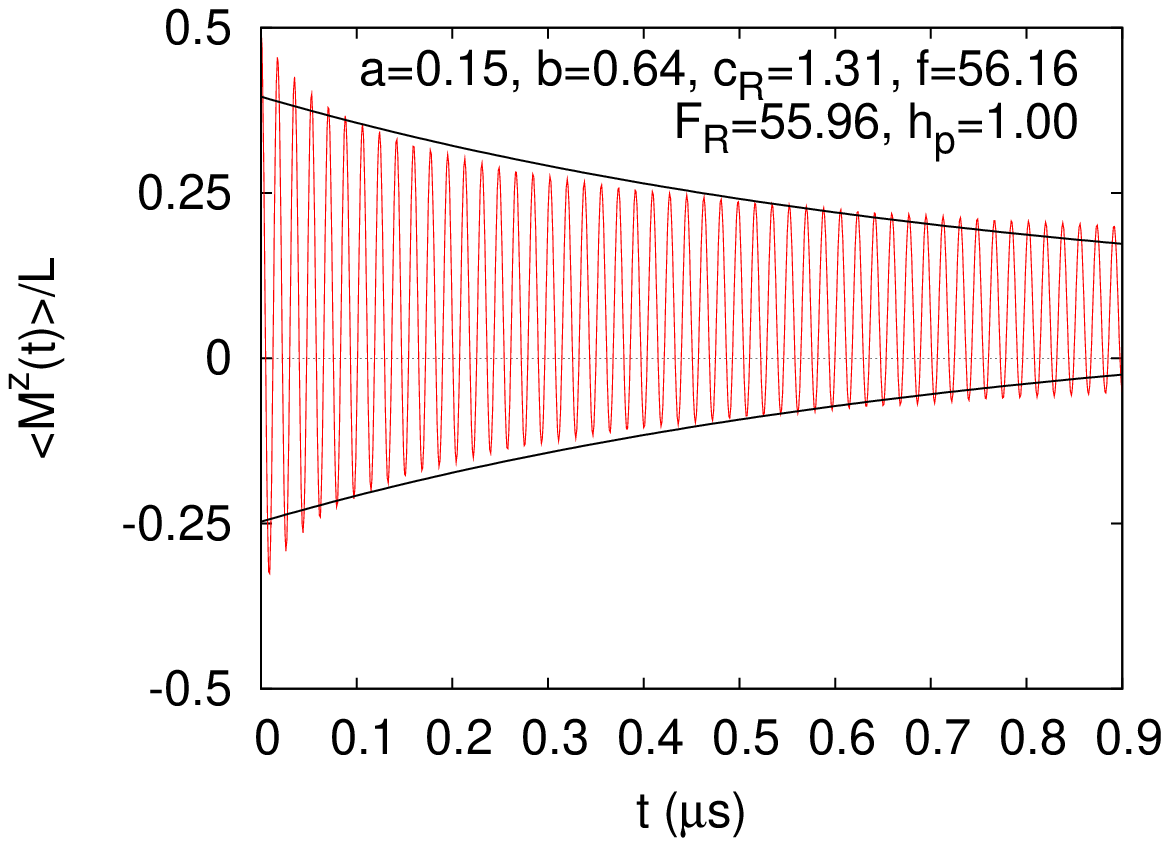}
\includegraphics[width=8cm]{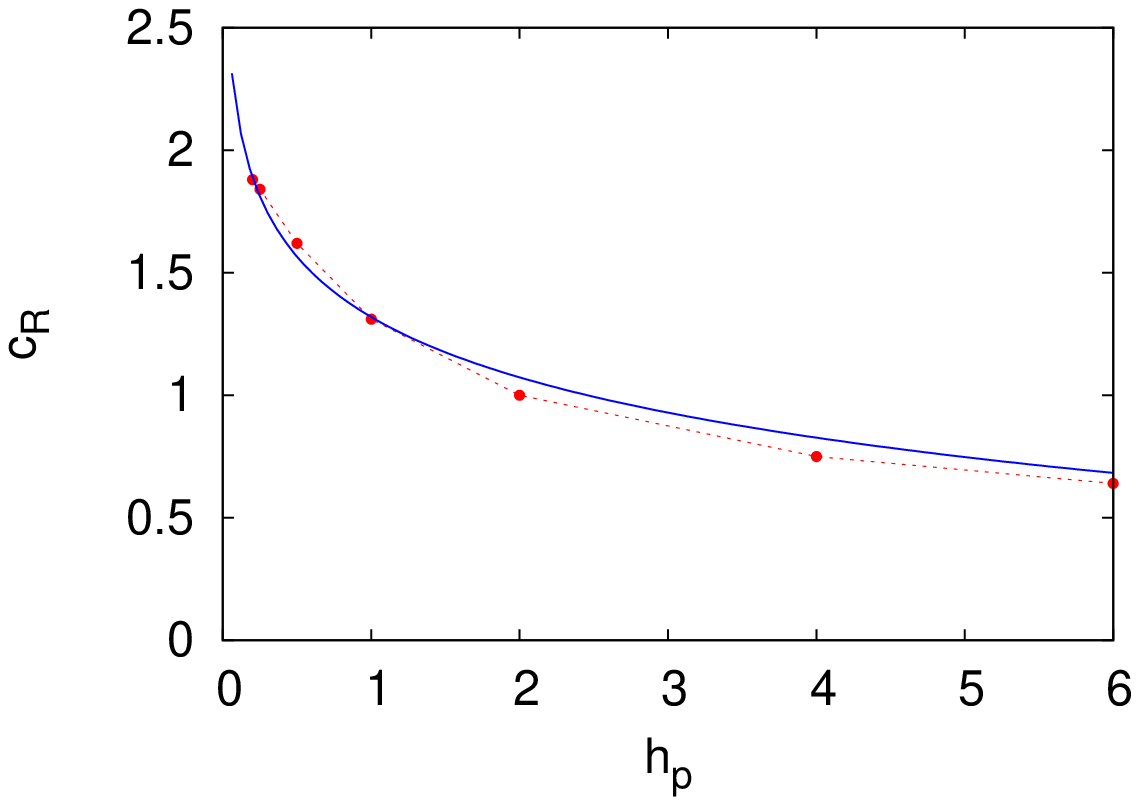}
\caption{(Color online)
Left:
Time evolution of the longitudinal magnetization as obtained by numerical evaluation
of Eq.~(\ref{Eq.Ca1.1}) (or the solution of the TDSE for $D_0=\xi^x=\xi^y=\zeta=0$ for $\Gamma=0.001$),
that is for the case that there are random fluctuations in the $g_z$-factor only.
The solid line represents the envelope $(a\pm be^{-c_\mathrm{R}t})/2$ of the function $(a+be^{-c_\mathrm{R}t}\cos2\pi ft)/2$ that was fitted to the data.
Right: The decay rate $c_\mathrm{R}$ as a function of $h_p$, obtained by fitting
$(a+be^{-c_\mathrm{R}t}\cos2\pi ft)/2$ to the time-dependent data.
The solid line shows that the function $a'\ln(h_p)+b'$ with $a'=-0.35$ and $b'=1.32$ fits the data reasonably well.
The dashed line connecting the bullets is a guide to the eye only.
}
\label{fig.Ca1}
\end{figure}

In Fig.~\ref{fig.Ca1}(left), we present a typical result for the time dependence of the
longitudinal magnetization with $g_z$-factor distribution (only) suggesting
that the time-averaged longitudinal magnetization is non-zero,
in concert with the analytical expressions
\begin{eqnarray}
\lim_{T\rightarrow\infty}\frac{1}{T}\int_0^T\overline{\langle S^x(t)\rangle}\,dt
&=&
0
,
\nonumber \\
\lim_{T\rightarrow\infty}\frac{1}{T}\int_0^T\overline{\langle S^y(t)\rangle}\,dt
&=&
0
,
\nonumber \\
\lim_{T\rightarrow\infty}\frac{1}{T}\int_0^T\overline{\langle S^z(t)\rangle}\,dt
&=&
\frac{1}{2}\frac{\Gamma F_0}{h_pF_\mathrm{R}+\Gamma F_0}
.
\label{Eq.Ca1.5}
\end{eqnarray}
The reason for this positive offset is simple:
Any non-zero field in the $z$-direction tilts the plane of the Rabi oscillations
away from the $(y,z)$-plane,
introducing a small precession about the tilted axis superimposed on the Rabi nutation,
leading to a positive long-time average.
This non-zero offset effect is significant because, as we will see later, it is a unique
signature of the presence of random fluctuations in the $g_z$ factor or,
equivalently, of the inhomogeneity of the static magnetic field.
We emphasize that this non-zero offset is due to randomness
and not due to dissipation, as the present paper considers the case
of $T_1=\infty$ only.

Similarly, in the case that all spins are initially along the $x$-direction,
the long-time average of the transverse magnetization is given by
\begin{eqnarray}
\lim_{T\rightarrow\infty}\frac{1}{T}\int_0^T\overline{\langle S^x(t)\rangle}\,dt
&=&
\frac{1}{2}\frac{h_pF_\mathrm{R}}{h_pF_\mathrm{R}+\Gamma F_0}
,
\label{Eq.Ca1.6}
\end{eqnarray}
the long-time averages of the two other components being zero.
Unlike in the case of the longitudinal magnetization,
in the regime where the transverse magnetization shows oscillations ($h_pF_\mathrm{R}>\Gamma F_0$),
the transverse magnetization reaches its asymptotic value Eq.~(\ref{Eq.Ca1.6}) already after a few
oscillations (data not shown).

From Eq.~(\ref{Eq.Ca1.1}), it is clear that we cannot expect the amplitude of
the Rabi oscillations to decay exponentially in a strict sense.
Nevertheless, the data fits well to a function of the form $(a+be^{-c_\mathrm{R}t}\cos2\pi ft)/2$.
The decay rate $c_\mathrm{R}$, shown in Fig.~\ref{fig.Ca1}(right),
{\bf decreases} with increasing microwave amplitude $h_p$.
It seems to diverge when $h_p\rightarrow0$ but this is never observed in experiment.

This decrease is a second characteristic feature of the presence of random fluctuations in the $g_z$ factor or,
equivalently, of the inhomogeneity of the static magnetic field.

\subsubsection{Randomness in $g_x$ and $g_y$: Non-interacting spins}\label{Ca2}

In this case, the Hamiltonian is given by Eq.~(\ref{M0}) with $\zeta_j=\xi^z=D_0=0$
and we have
\begin{eqnarray}
\overline{\langle S^z(t)\rangle}
&=&
\frac{1}{2} \int_{-\infty}^{+\infty} 
\cos \left[ \Omega_{\mathrm{R}} t(1+(\xi^x+\xi^y)/2) \right]
p(\xi^x)
p(\xi^y)
\,d\xi^x\,d\xi^y
.
\label{simp2c}
\end{eqnarray}
Taking the cut-off $\xi_0$ to be infinity we obtain
\begin{eqnarray}
\overline{\langle S^z(t)\rangle}
&=&
\frac{1}{2\pi} \int_{-\infty}^{+\infty} 
\cos \left[ \Omega_{\mathrm{R}} t(1+\zeta) \right]
\frac{\Gamma}{\zeta^2+\Gamma^2}
\,d\zeta
=\frac{1}{2} e^{-\Omega_{\mathrm{R}}\Gamma t} \cos \Omega_{\mathrm{R}} t
.
\label{simp2d}
\end{eqnarray}
Thus, we conclude that if there is randomness in $g_x$ and $g_y$ only,
the Rabi oscillations will decay exponentially with a rate proportional to $\Omega_R=2\pi h_pF_R$.
In the absence of the microwave field, the transverse magnetization
is a constant of motion and hence $T_2=\infty$.
Summarizing, in the presence of randomness in $g_x$ and $g_y$ only
and in the absence of dipole-dipole interactions, we have
\begin{equation}
1/T_{\mathrm{R}}=\Gamma \Omega_{\mathrm{R}} > 1/T_{\mathrm{2}}=0
,
\label{Eq.Ca.2}
\end{equation}
showing that the decay rate of the Rabi oscillations increases linearly
with the microwave amplitude $h_p$.
In fact, Eq.~(\ref{Eq.Ca.2}) is the same as Eq.~(\ref{Eq.Ba.2}) with
$\gamma$ replaced by $\Gamma$.
Thus, we conclude that randomness in $g_x$ and $g_y$ has
the same effect as randomness in the amplitude of the microwave field:
The Rabi oscillations decay exponentially, with a decay rate that
increases linearly with $\Omega_R=2\pi h_pF_R$.
In both cases, decoherence results from a loss of phase of superposed radiation emitted by spins in nutation leading,
as a consequence, to  energy transfer from the spin-bath to the electromagnetic bath.
Clearly enough such dissipation does not involve the usual relaxation time $T_1$ due to dissipation by phonons.
This case is very different from the one of e.g. superconducting qubits where
decoherence is dominated by $T_1$ process, as shown
for example in Ref.~\cite{SCHR08}.

\begin{figure}[t]
\includegraphics[width=8cm]{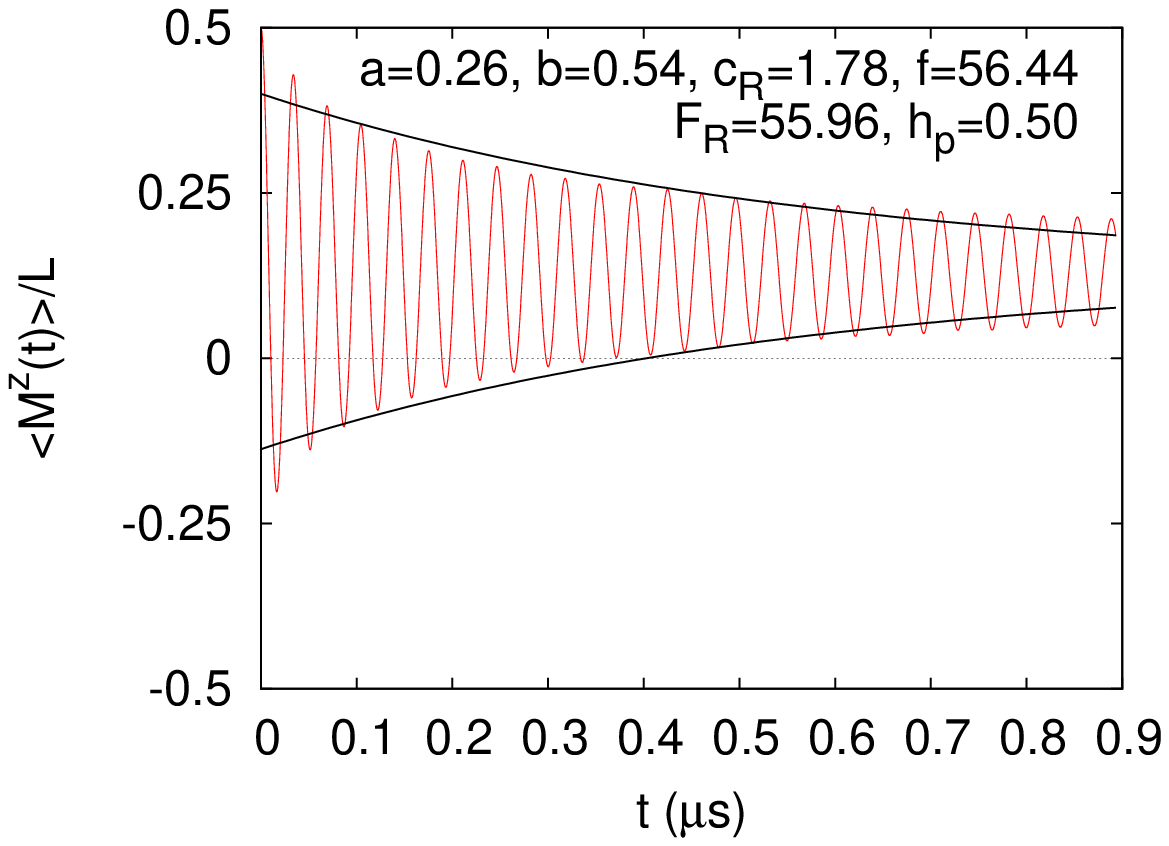}
\includegraphics[width=8cm]{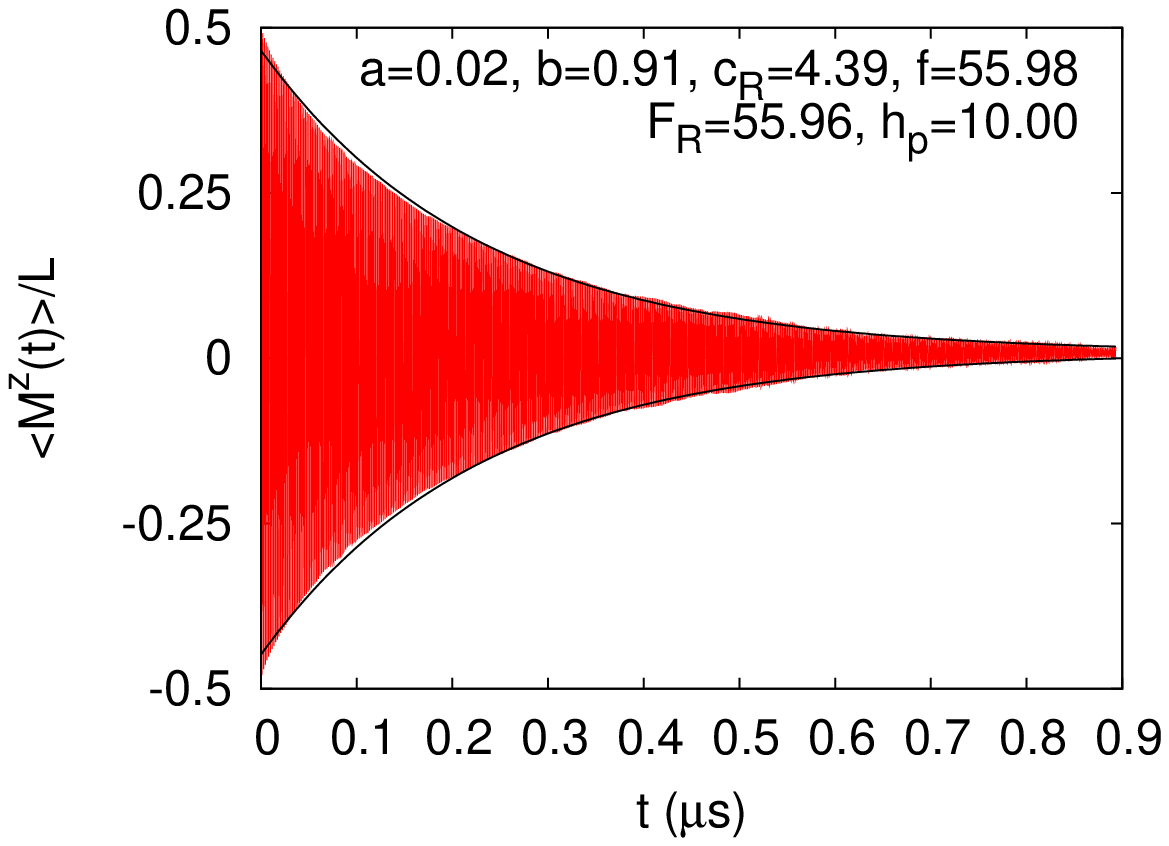}
\includegraphics[width=8cm]{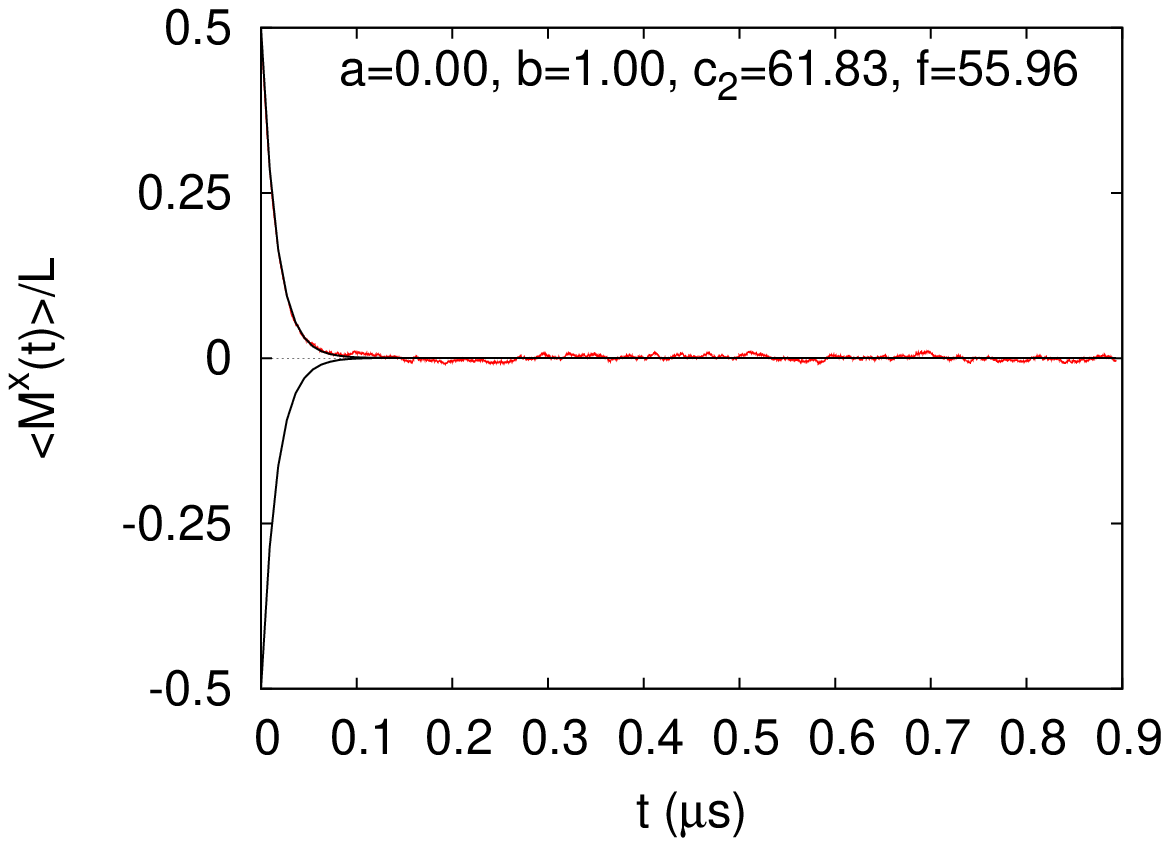}
\includegraphics[width=8cm]{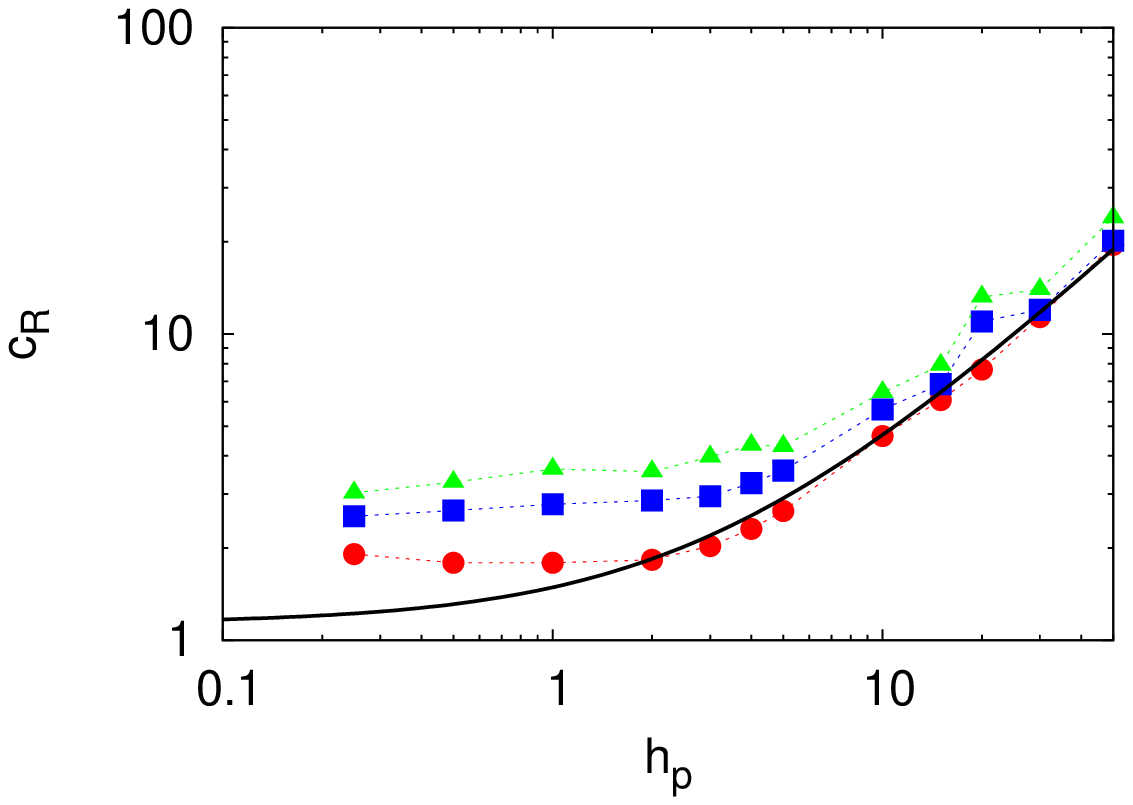}
\caption{(Color online)
Top left:
Time evolution of the longitudinal magnetization as obtained by numerical solution
of the TDSE (see Eq.~({\ref{TDSE}}))
for the case that there are random fluctuations in all three $g$-factors only
($\Gamma=\Gamma_x=\Gamma_y=\Gamma_z=0.001$ and $\zeta=D_0=0$).
The solid line represents the envelope $(a\pm be^{-c_\mathrm{R}t})/2$
of the function $(a+be^{-c_\mathrm{R}t}\cos2\pi ft)/2$ that was fitted to the data.
Top right: Same as top righ, except that $h_p=10$ instead of $h_p=0.5$.
Bottom left: Transverse magnetization in the absence of the microwave field ($h_p=0$).
The decay rate $c_2=60.19$ is in excellent agreement with the analytical
result $c_\mathrm{R}=2\pi\Gamma F_0=60.95$ predicted by Eq.~(\ref{Eq.Ca1.4}).
The solid line represents the function $(a+ be^{-c_2t})/2$ that was fitted to the data.
Bottom right: The inverse relaxation time $c_\mathrm{R}$ as a function of the microwave amplitude $h_p$
for $\Gamma_x=\Gamma_y=\Gamma_x=0.001$ and $\Gamma_z=0.001$ (bullets), $\Gamma_z=0.002$ (squares), $\Gamma_z=0.003$ (triangles).
The solid line represents the linear fit to the $\Gamma_z=0.001$ data.
The dashed lines are guides to the eye only.
The number of spins in these calculations is 10000.
}
\label{fig.Ca3.1}
\end{figure}

\subsubsection{Randomness in $g_x$, $g_y$ and $g_z$: Non-interacting spins}\label{Ca3}

In this case, the Hamiltonian is given by Eq.~(\ref{M0}) with $\zeta_j=D_0=0$.
In Fig.~\ref{fig.Ca3.1}(top),
we present a typical result for the time dependence of the longitudinal magnetization.
It is seen that the time-averaged longitudinal magnetization is non-zero,
signaling the presence of fluctuations in $g_z$ (see Section~\ref{Ca1}).
Also clearly visible is the increase of the decay rate $c_\mathrm{R}$
of the Rabi oscillations with increasing microwave amplitude $h_p$,
a signal of the presence of fluctuations in $(g_x,g_y)$ (see Section~\ref{Ca2}).
Note that there is no obvious relation between the decay rate
of the transverse magnetization ($c_2\approx 60$, see Fig.~\ref{fig.Ca3.1}(bottom left)) and
the values of the decay rate $c_\mathrm{R}$ at the smallest values
of $h_p$ shown in Fig.~\ref{fig.Ca3.1}(bottom right).

From the results of Sections~\ref{Ca1} and \ref{Ca2}, we may expect that
the decay rate $c_\mathrm{R}$ shows a crossover from the  regime
in which the fluctuations on $g_z$ dominate ($c_\mathrm{R}$ decreases
with increasing $h_p$) and a
regime in which the fluctuations on $(g_x,g_y)$ dominate ($c_\mathrm{R}$ increases
linearly with $h_p$).
This is borne out by the data presented in Fig.~{\ref{fig.Ca3.1}}(bottom right)
where we show the combined effect of the two different
sources of decoherence, the widths
of the Lorenztian distributions for the longitudinal ($g_z$, $\Gamma_z$) and transverse (
$(g_x,g_y)$, $\Gamma_x=\Gamma_y$) fluctuations being varied independently.

\begin{figure}[t]
\includegraphics[width=12cm]{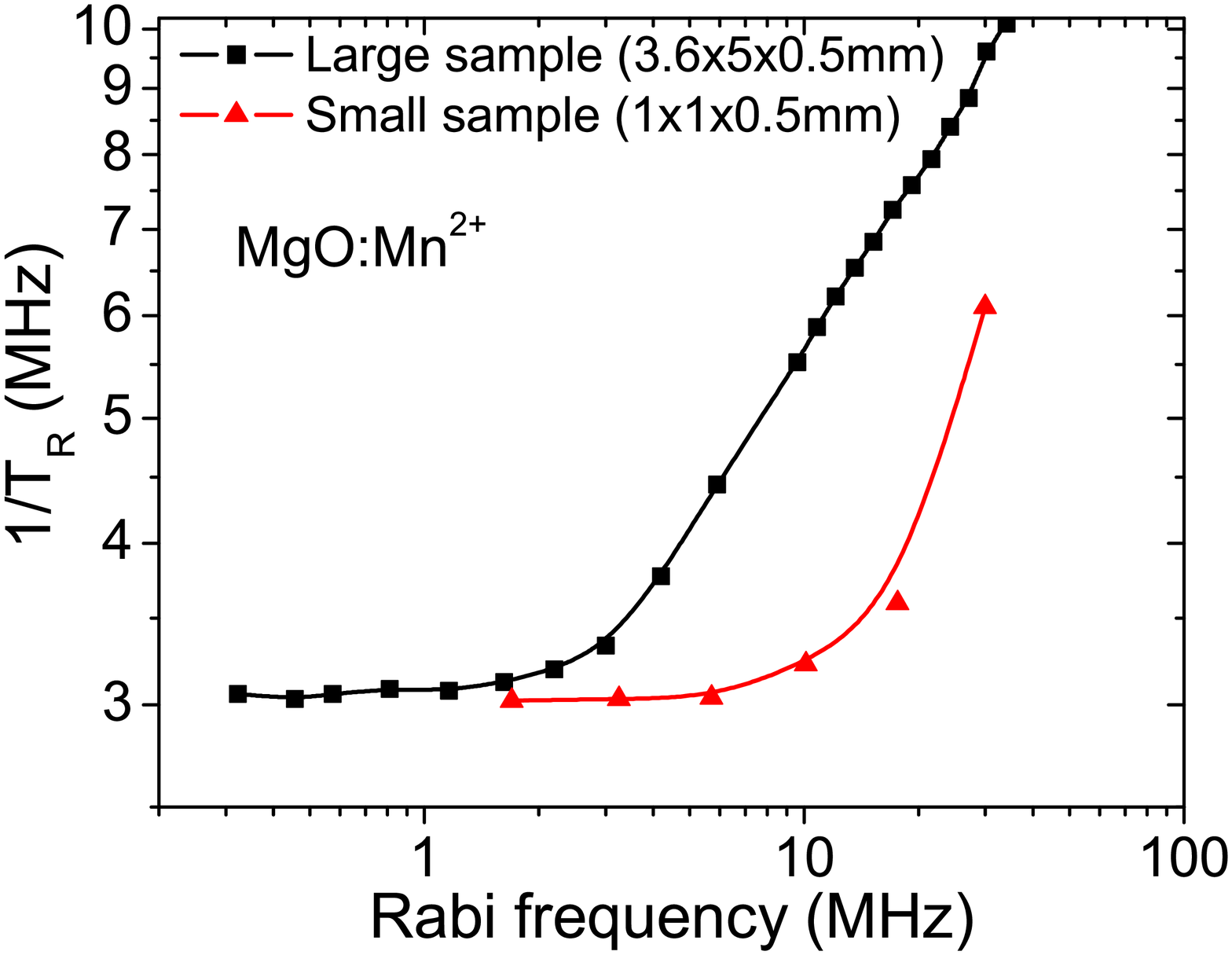}
\caption{(Color online) Decay time of the Rabi oscillations of MgO:Mn$^{2+}$ (0.001\%)
as a function of the microwave field amplitude (Rabi frequency $\Omega_R$)
for two samples of different sizes.
Measurements were carried out at room temperature.
}
\label{fig:Mn2}
\end{figure}

\subsubsection{Experimental results: MgO:Mn$^{2+}$}

The combined effect of a distribution in the $g$-factors and inhomogeneities in
the microwave amplitude
are shown in experiments performed on single crystalline films of MgO:Mn$^{2+}$, see Fig.~\ref{fig:Mn2}
where the measured Rabi dacay time is plotted versus the Rabi frequency.
The Mn$^{2+}$ dilution is such that dipolar interactions are negligible.
Due to weak but sizable distributions of Mn$^{2+}$ local environments,
we expect non-negligible and similar distributions of the three $g$-factor components.
For small microwave amplitudes, the
distribution in the $g_z$-factor gives the dominant, nearly constant contribution
to the Rabi decay time, which compares well with Fig.~\ref{fig.Ca3.1}(bottom right).
As the microwave amplitude increases,
the inhomogeneities associated with transverse components take over
and $1/\tauR $ increases linearly on the log-log scale.
Note that the slope of one-half differs from the slope one that we have for the model considered in this paper.
This is because of the peculiarity of the experimental system
where nutation takes place coherently over five equidistant levels of the material, an aspect
that will be considered in the future.
At present, we are interested in showing that the departure
from the $1/\tauR $ plateau takes place more rapidly
with the larger sample as expected when the effect of microwave inhomogeneities
dominates over the one of $g$-factor distributions.

\begin{figure}[t]
\includegraphics[width=8cm]{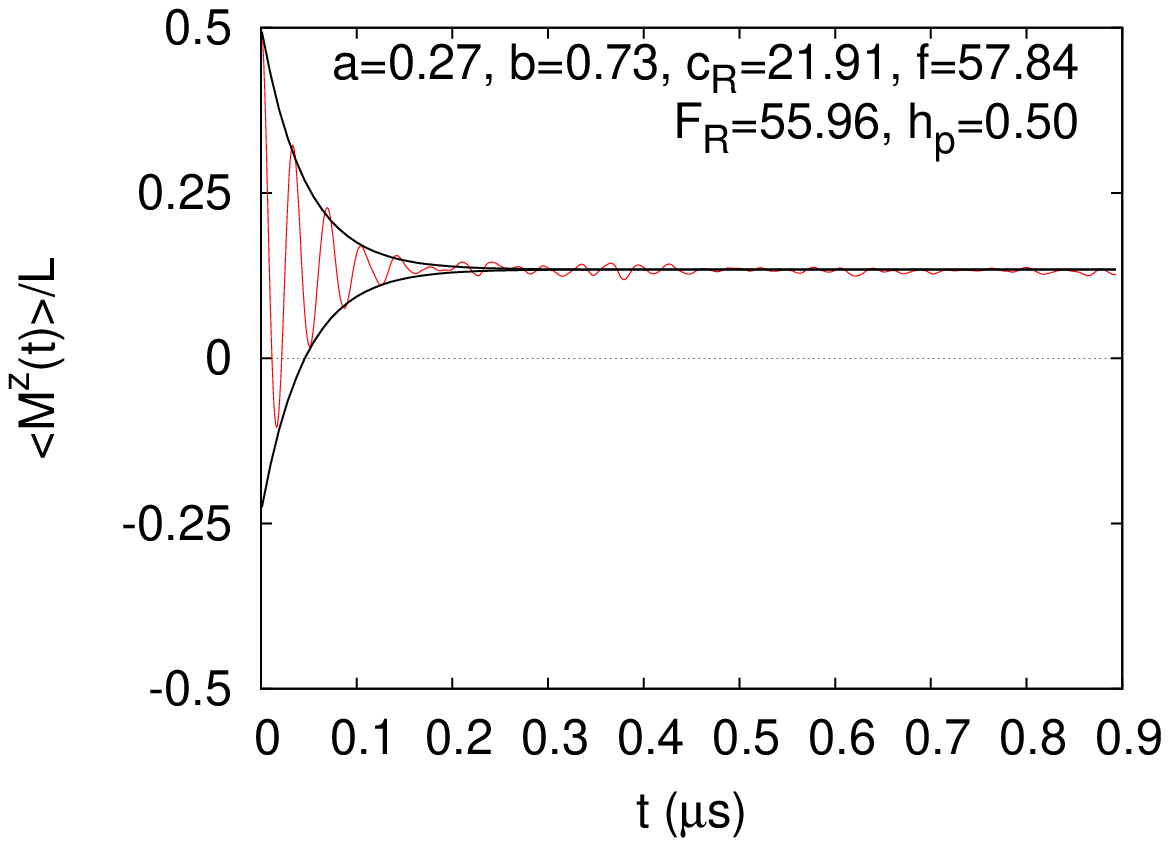}
\includegraphics[width=8cm]{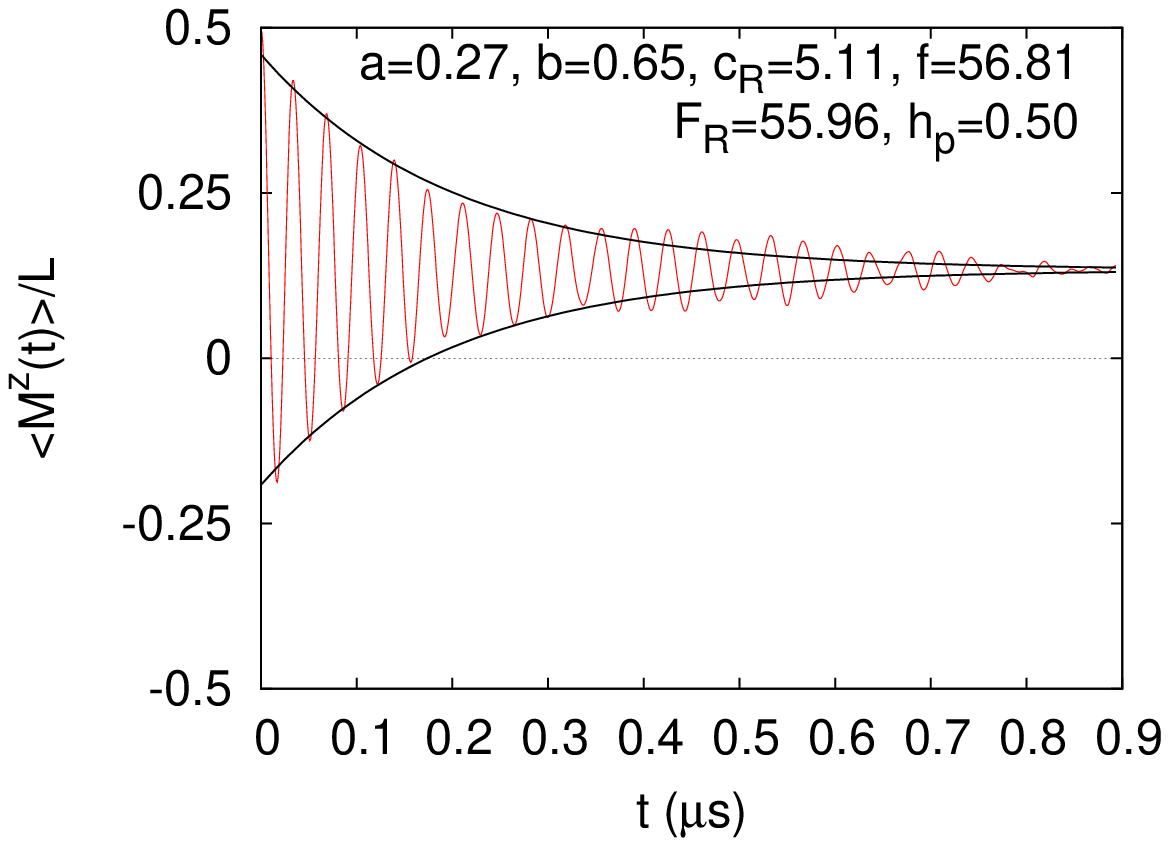}
\includegraphics[width=8cm]{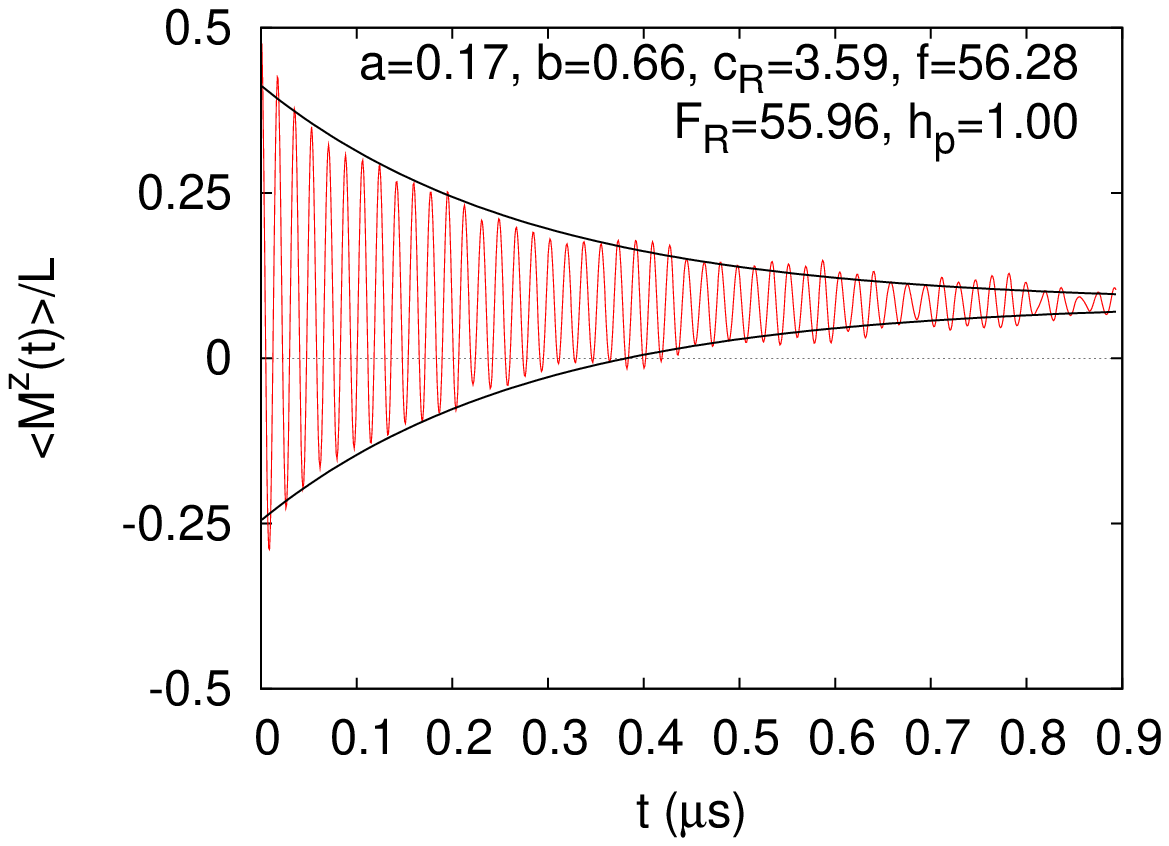}
\includegraphics[width=8cm]{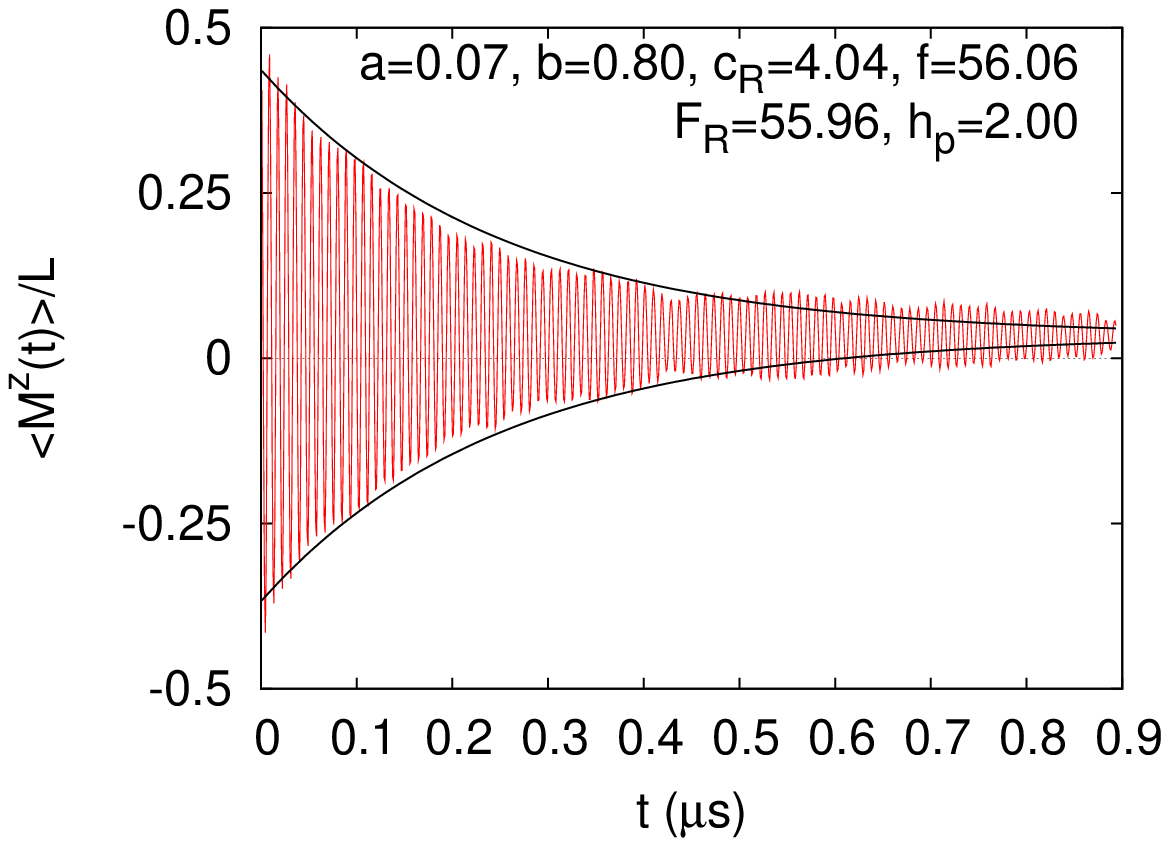}
\includegraphics[width=8cm]{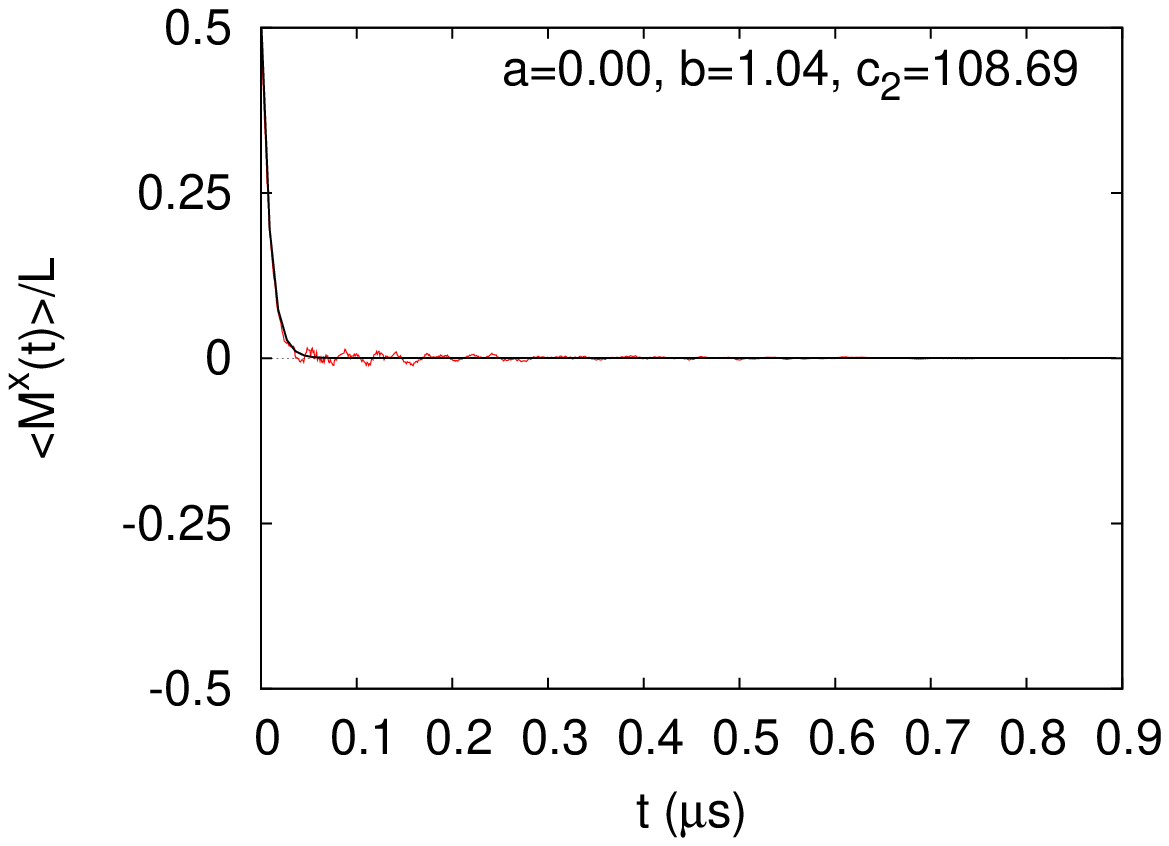}
\includegraphics[width=8cm]{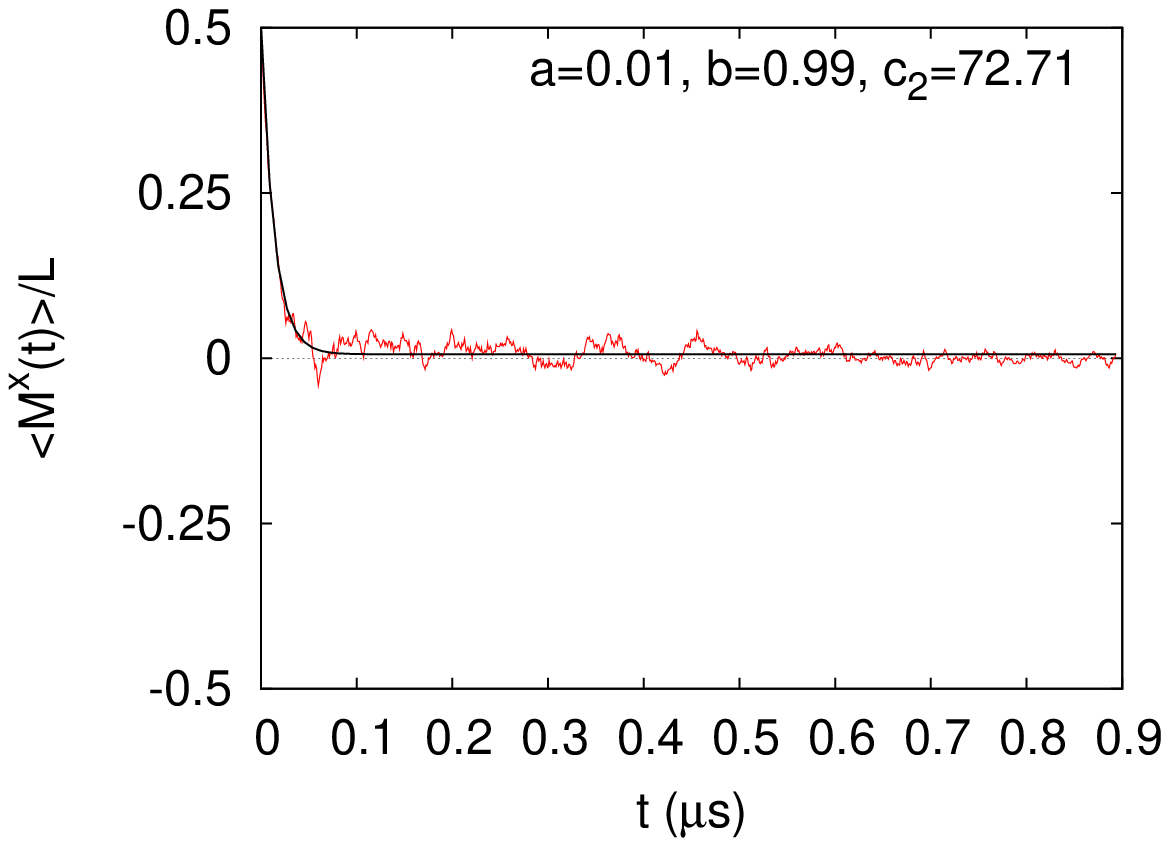}
\caption{(Color online)
The Rabi oscillations of the longitudinal magnetization as obtained by solving the TDSE
with the Hamiltonian Eq.~(\ref{M0}) for 26 spins that interact via dipole-dipole interaction,
for different concentrations $\concentration$,
with random fluctuations in the three $g$-factors ($\Gamma=0.001$) and
without random fluctuations in the microwave amplitude ($\gamma=0$).
Top left: $\concentration=10^{-3}$;
Top right to middle right: $\concentration=10^{-4}$.
The solid line represents the envelope $(a\pm be^{-c_\mathrm{R}t})/2$ of the function $(a+be^{-c_\mathrm{R}t}\cos2\pi ft)/2$ that was fitted to the data.
Bottom: Time evolution of the transverse magnetization for $\concentration=10^{-3}$ (left)
and $\concentration=10^{-4}$ (right).
The solid line represents the function $(a+ be^{-c_2t})/2$ that was fitted to the data.
}
\label{fig.Cb.1}
\end{figure}

\begin{figure}[t]
\includegraphics[width=8cm]{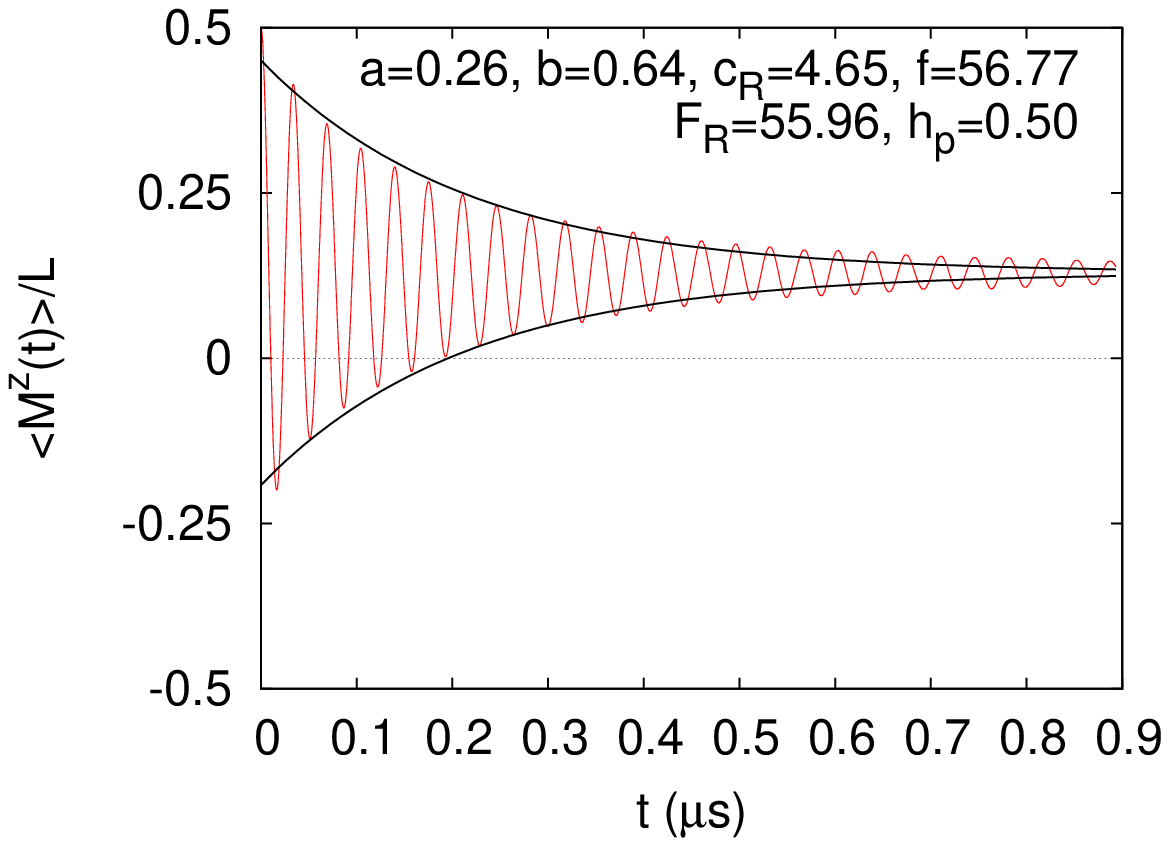}
\includegraphics[width=8cm]{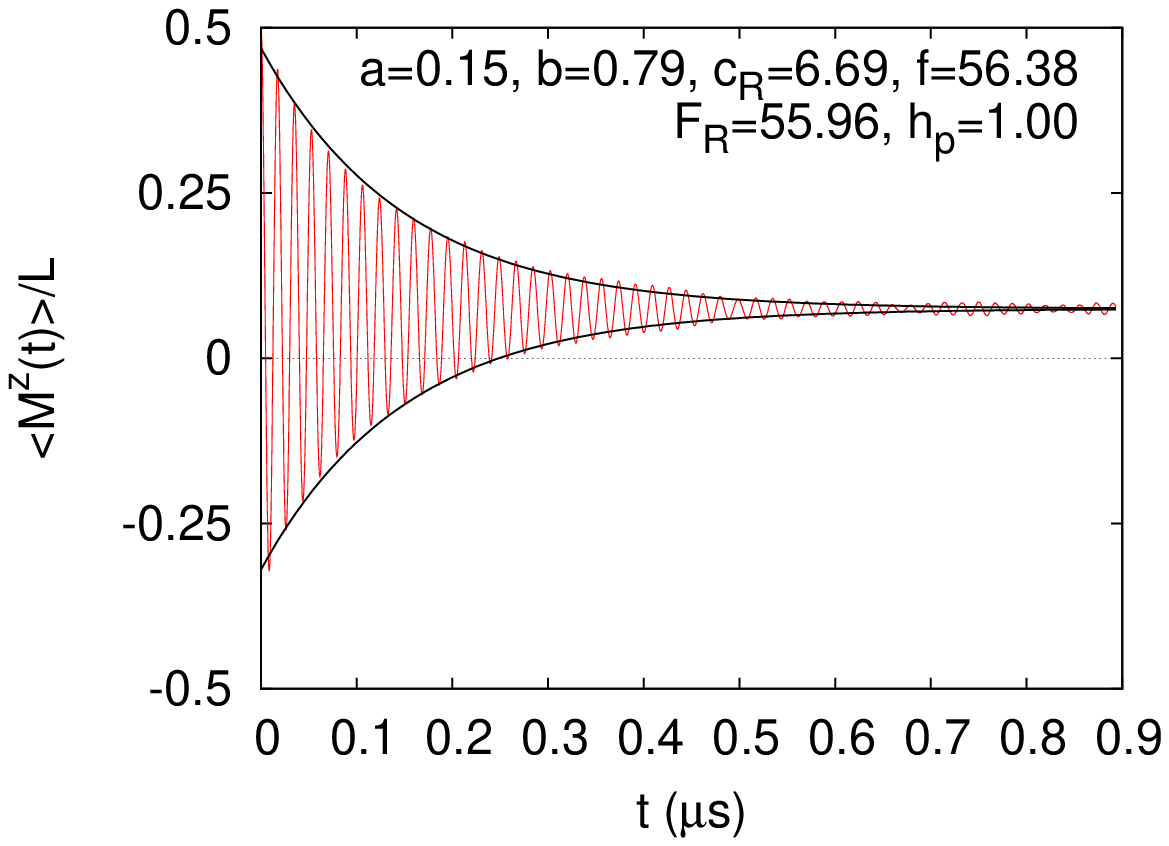}
\includegraphics[width=8cm]{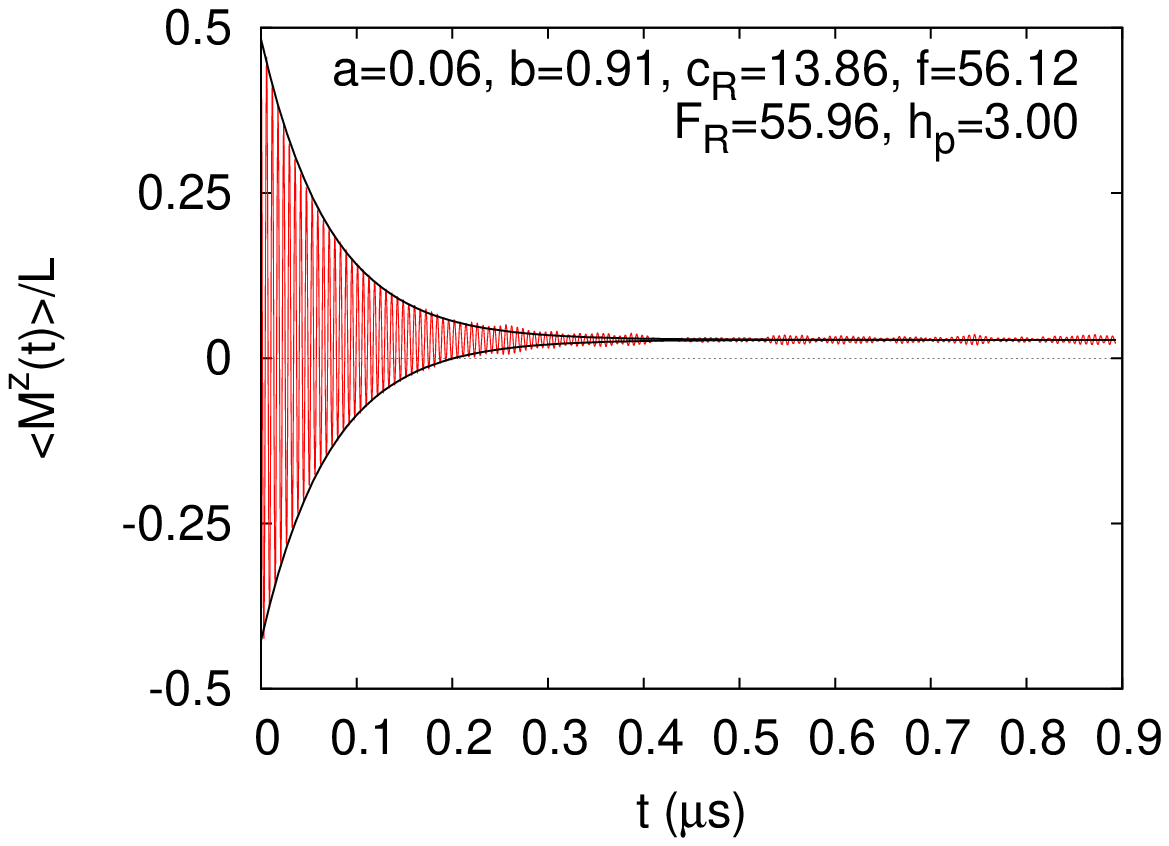}
\includegraphics[width=8cm]{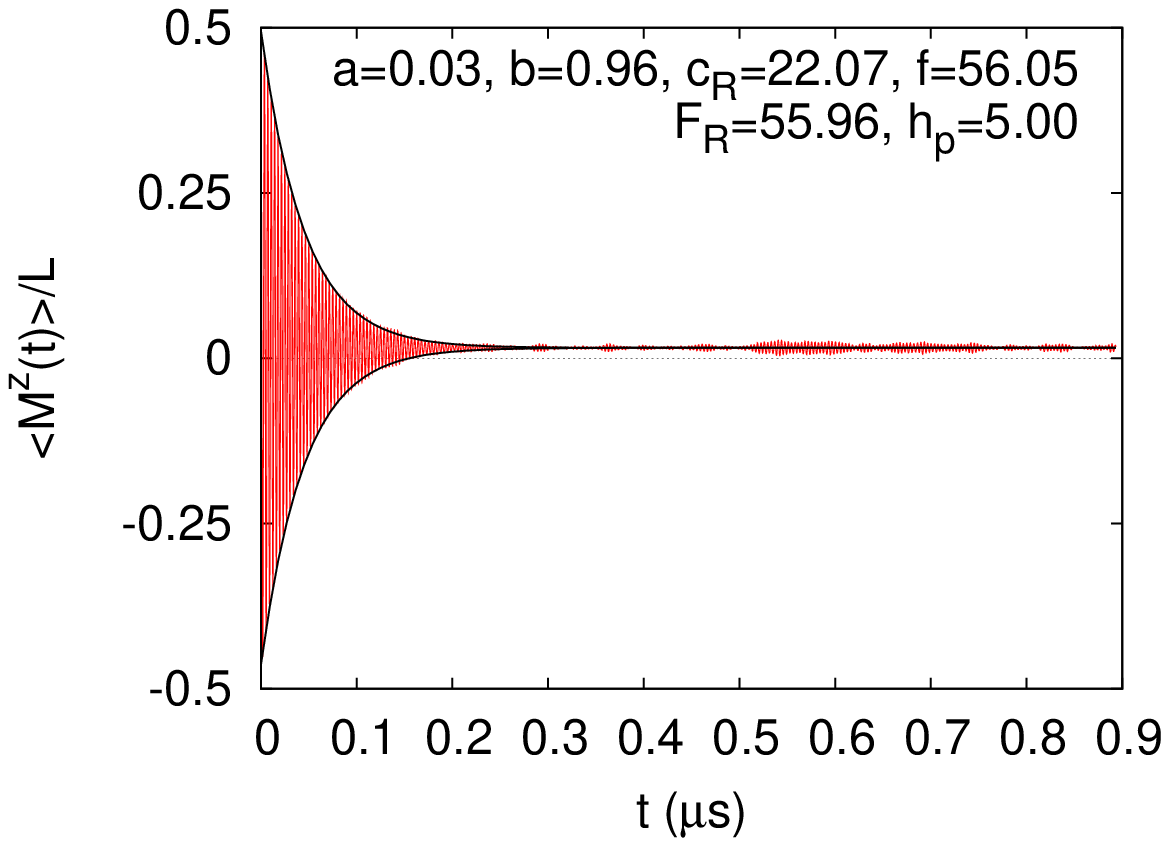}
\caption{(Color online)
The Rabi oscillations as obtained from the numerical solution
of the TDSE (see Eq.~({\ref{TDSE}})) for $D_0=0$, $\gamma=0.01$ and $\Gamma=0.001$,
that is for the case that there are random fluctuations
in both the microwave field and in the $g$-factors.
The solid line represents the envelope $(a\pm be^{-c_\mathrm{R}t})/2$ of the function $(a+be^{-c_\mathrm{R}t}\cos2\pi ft)/2$ that was fitted to the data.
The number of spins in these calculations is $10000$.
}
\label{fig.Da.1}
\end{figure}

\subsubsection{Dipolar-coupled spins}\label{Cb}

In Fig.~\ref{fig.Cb.1}(top and middle), we present simulation results for systems of 26 spins with dipole-dipole interaction,
(with different concentrations $n$), with random fluctuations in the three $g$-factors and uniform microwave field amplitude.
These results are obtained by averaging over ten different realizations, meaning ten different distributions of the 26 spins on the lattice.
The striking signature of the presence of fluctuations in $g_z$, namely the non-zero long-time average of the
longitudinal magnetization, remains untouched by the effects of the dipolar interactions.
For the values of $h_p$ shown in Fig.~\ref{fig.Cb.1}(top left to middle right), the dependence of the decay rate $c_\mathrm{R}$
is essentially the same as if the dipolar interactions were absent (see Fig~\ref{fig.Ca3.1}(bottom right)).
For large $h_p$ (data not shown), the decay rate $c_\mathrm{R}$ linearly increases with $h_p$.
Comparing Fig.~\ref{fig.Cb.1}(bottom left) with Fig.~\ref{fig.Cb.1}(bottom right), it follows that the value of the
decay rate of the transverse magnetization is nearly independent of the concentration,
hence cannot be attributed to the presence of dipolar interactions but is
mainly due to the presence of fluctuations in $g_z$.

\begin{figure}[t]
\includegraphics[width=8cm]{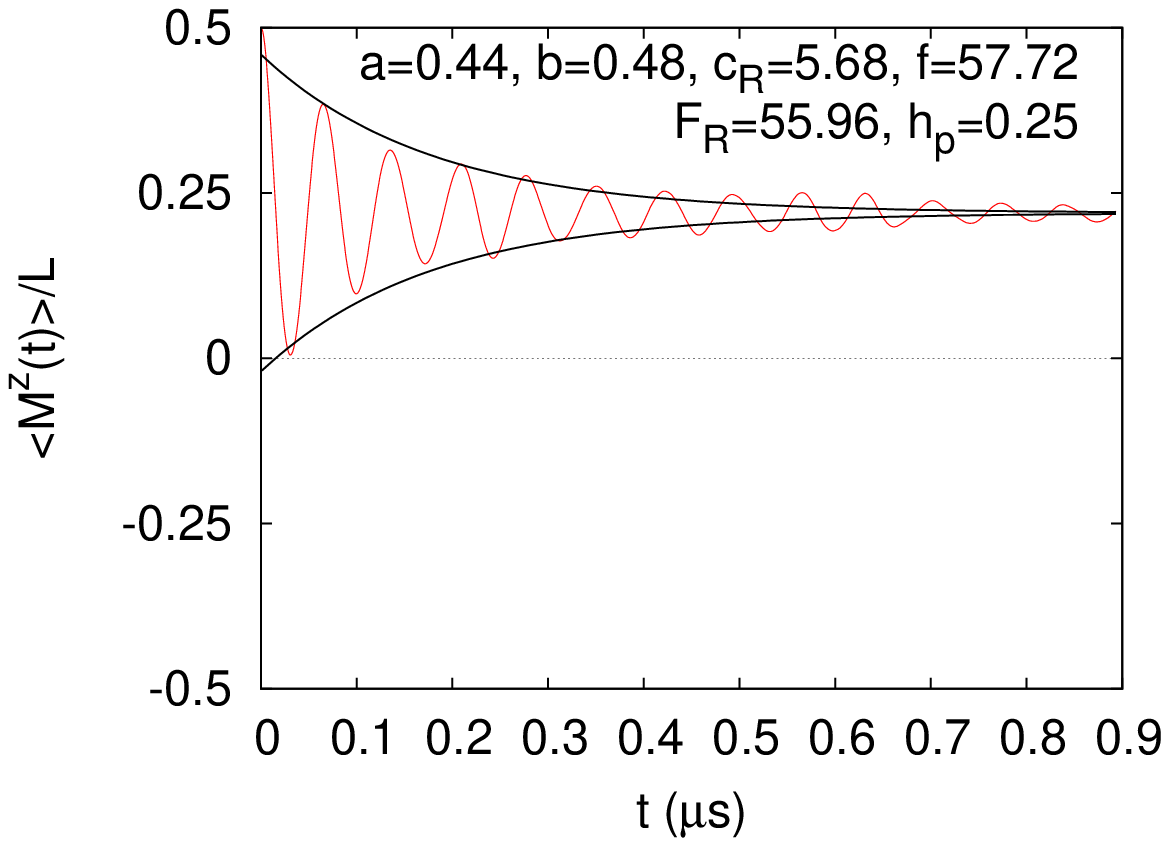}
\includegraphics[width=8cm]{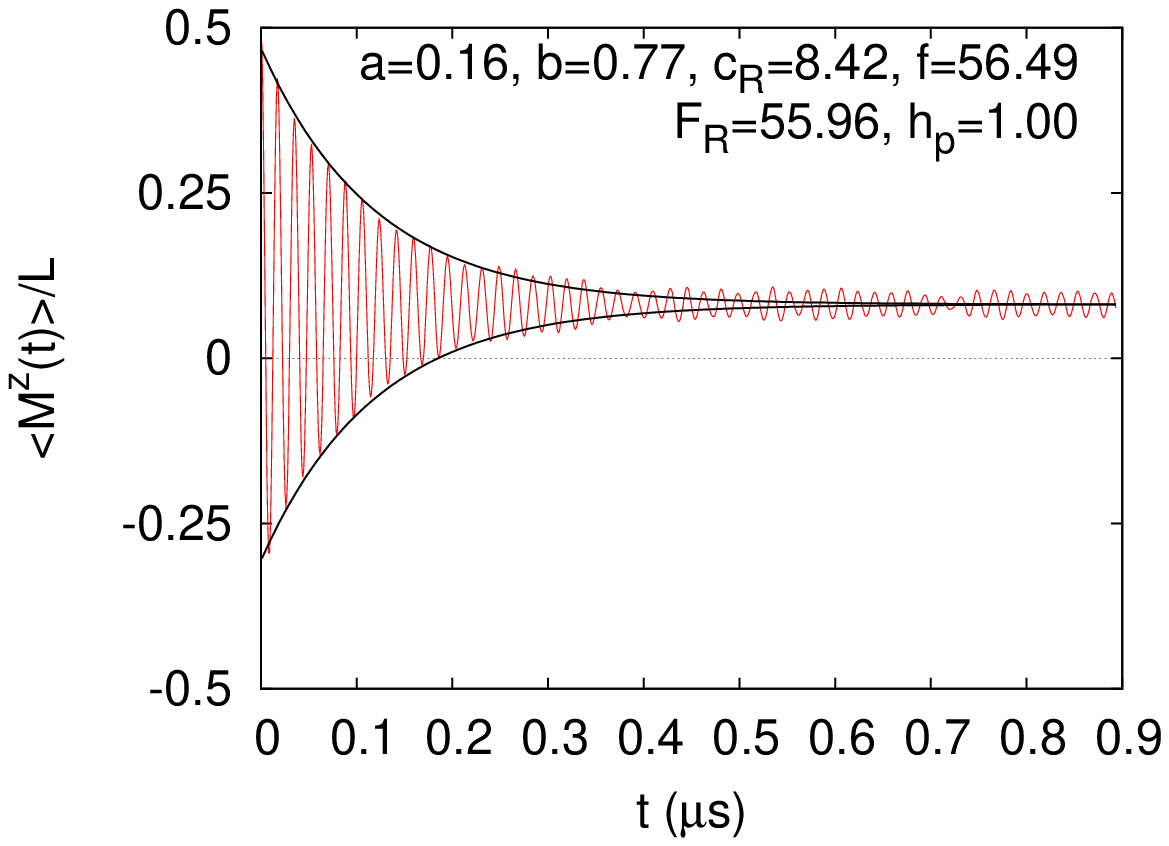}
\includegraphics[width=8cm]{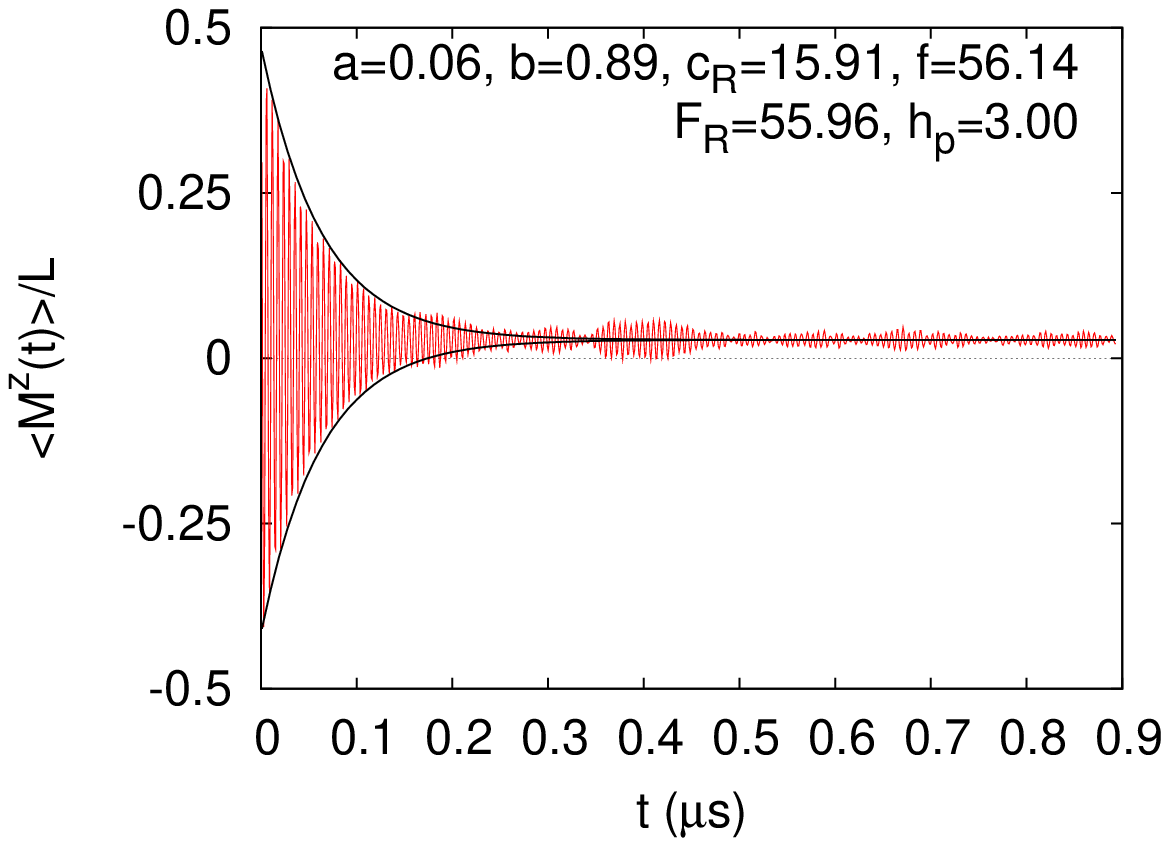}
\includegraphics[width=8cm]{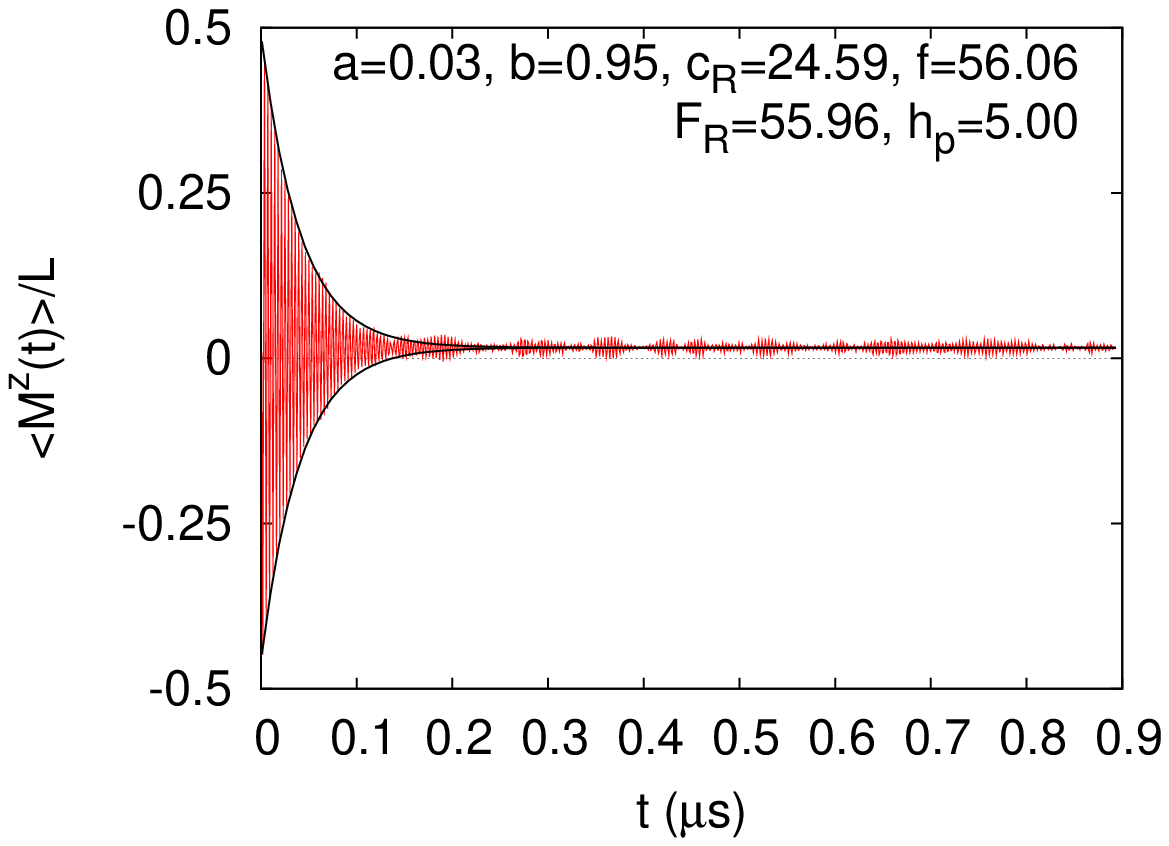}
\includegraphics[width=8cm]{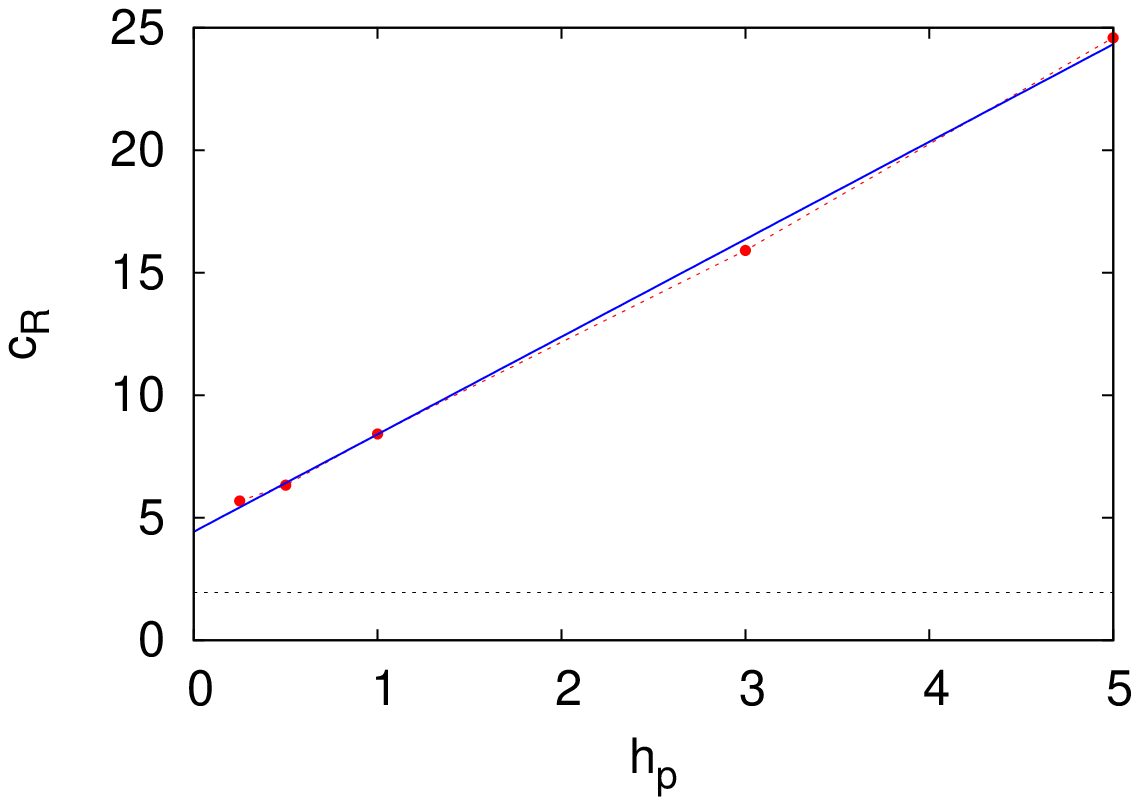}
\includegraphics[width=8cm]{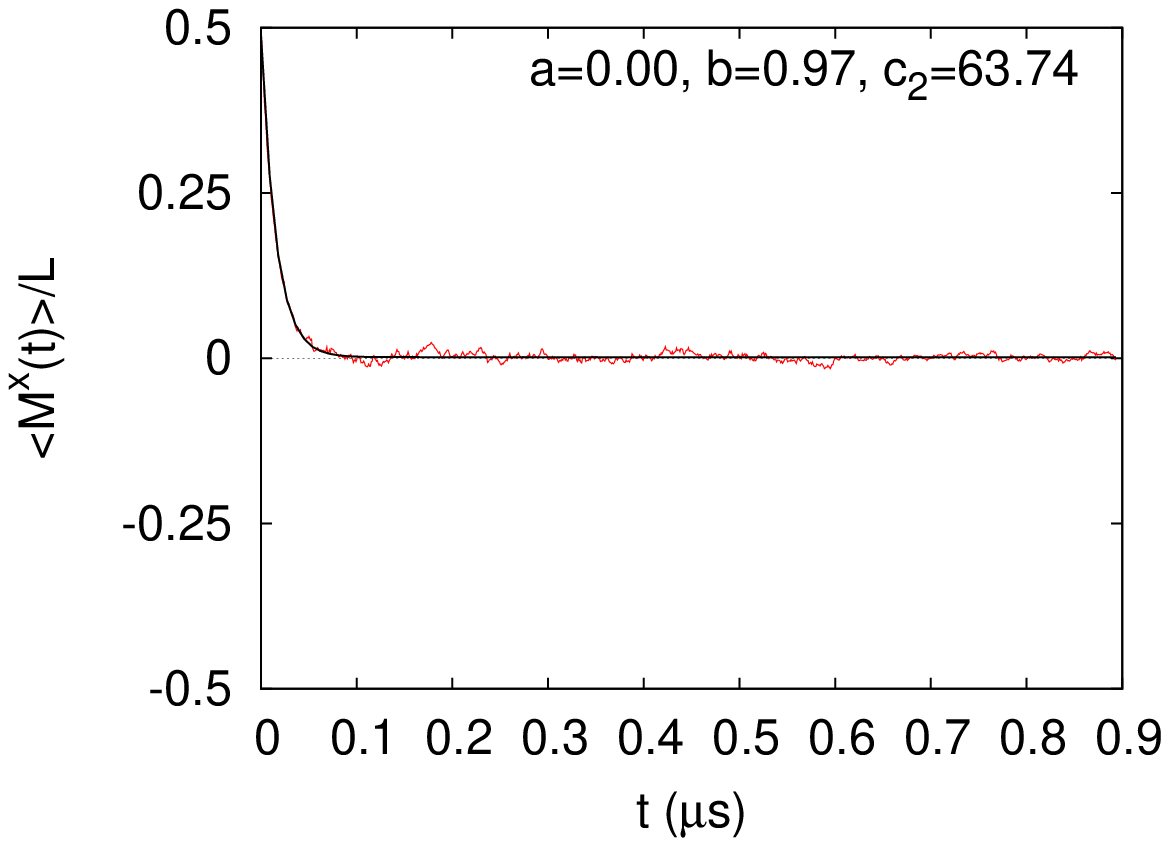}
\caption{(Color online)
Simulation results as obtained by solving the TDSE
for the Hamiltonian Eq.~(\ref{M0}) for 12 spins (concentration $\concentration=10^{-4}$)
that interact via dipole-dipole coupling,
with random fluctuations in the $g$-factors ($\Gamma=0.001$),
and with random fluctuations in the microwave amplitude ($\gamma=0.01$).
Top left to middle right: Longitudinal magnetization showing Rabi oscillations.
The solid line represents the envelope $(a\pm be^{-c_\mathrm{R}t})/2$ of the function
$(a+be^{-c_\mathrm{R}t}\cos2\pi ft)/2$ that was fitted to the data.
Bottom left:
Bullets show the inverse relaxation time $c_\mathrm{R}=1/T_\mathrm{R}$ as a function of the microwave amplitude $h_p$.
The dashed line connecting the bullets is a guide to the eye only.
A linear fit to the simulation data yields $c_\mathrm{R}=1/T_\mathrm{R}\approx 3.98 h_p + 4.43$ and is shown by the solid line.
The horizontal line represents the value of $1/2T_2\approx 1.95$, estimated from the data
of the transverse magnetization in the absence of random fluctuations in the $g$-factors
and on the microwave amplitude (see Fig.~\ref{fig.Bb}).
Bottom right:
Transverse magnetization in the absence of the microwave field ($h_p=0$).
The solid line represents the function $(a+ be^{-c_2t})/2$ that was fitted to the data.
The decay rate $c_2$ contains contributions from the dipolar interactions and,
most importantly, from the random fluctuations in $g_z$.
}
\label{fig.Db.1}
\end{figure}

\subsection{Randomness in the $g$-factors and the microwave amplitude}\label{D}

\subsubsection{Non-interacting spins}\label{Da}
In Fig.~\ref{fig.Da.1}, we present a few representative results
for the case that there are random fluctuations
in both the microwave amplitude and in the $g$-factors,
as obtained by solving the TDSE for the Hamiltonian Eq.~(\ref{M0}) with $D_0=0$.
In essence, the results are very similar to those of the case
where there are fluctuations in all three $g$-factors only.
This is easy to understand from Eq.~(\ref{M0}): Fluctuations in $(g_x,g_y)$ or (exclusive)
in the microwave amplitude have the same effect on the decay of the Rabi oscillations.
With both types of fluctuations present, our numerical results show that this contribution
does not significantly alter the dependence of $c_\mathrm{R}$ on $h_p$.

As before, the presence of fluctuations in $g_z$ (see Section~\ref{Ca1})
is signaled by the time-averaged longitudinal magnetization being non-zero
and by a contribution to the decay rate $c_2$ of the transverse magnetization,
which is in excellent agreement with the analytical
result $c_2=2\pi\Gamma F_0$ predicted by Eq.~(\ref{Eq.Ca1.4}) (data not shown).
Thus, in this case, we obviously have $c_2 >c_\mathrm{R}$ which is the same as $T_R > T_2$
where $T_2$ is reduced by the fluctuations in $g_z$.

\subsubsection{Dipolar-coupled spins}\label{Db}

In Fig.~\ref{fig.Db.1}, we present simulation results for systems of 12 spins with dipole-dipole interaction,
as obtained by averaging the solution of the TDSE over 100 different distributions of the 12 spins on the lattice,
for the case that there are random fluctuations in the microwave amplitude and in all three $g$-factors.

The four upper panels of Fig.~\ref{fig.Db.1} show results for the longitudinal magnetization.
The decay of the longitudinal magnetization is exponential to good approximation.
The signature of the presence of fluctuations in $g_z$, namely the non-zero long-time average of the
longitudinal magnetization is clearly visible.
For the values of $h_p$ shown in Fig.~\ref{fig.Db.1}(bottom left), the linear dependence of the decay rate $c_\mathrm{R}$
is essentially the same as if the dipolar interactions were absent (see Fig~\ref{fig.Ca3.1}(bottom right)).

A linear fit to the data of $c_\mathrm{R}$ yields $\lim_{h_p\rightarrow0}c_\mathrm{R}\approx 4.43$.
This value should be contrasted with the result $c_2\approx63.74$
for the transverse magnetization in the absence of microwaves ($h_p=0$) (see Fig.~\ref{fig.Db.1}(bottom right)).
Such a large $c_2$ (small $T_2$) resulting from both
dipolar interactions and fluctuations on all $g$-factors is effectively caused by the effect of $g_z$-fluctuations,
in concert with the results shown in Fig.~\ref{fig.Cb.1}(bottom) that
demonstrate that the concentration dependence is weak,
implying that the effect of the dipolar interactions is small
compared to that of the presence of fluctuations in $g_z$.

According to theory, the total decay rate of the transverse magnetization
is the sum of the decay rates due to the dipolar interactions only and the combined decay rate
due to field inhomogeneities only. From Fig.~\ref{fig.Bb}, the former is given by $c_2\approx3.90$.
In the absence of dipolar interactions, the latter is given by $c_2=2\pi\Gamma F_0=60.95$ (see Section~\ref{Ca3},
and Fig.~\ref{fig.Ca3.1},  yielding $c_2\approx60.19$ for $\gamma=0$).
Therefore, we have $c_2^{total}\approx64.85$, in very good agreement with the value $c_2=63.74$ extracted from
the simulation (see bottom right panel of Fig.~\ref{fig.Db.1}).

\section{Phenomenological model}\label{BE}

The simulations of the dipolar-coupled spin systems are rather expensive in terms of computational
resources. For instance, one simulation of a single realization of a 26-spin system takes about 20 hours, using 512 CPUs
on an IBM BlueGene/P.
Such relatively expensive simulations are necessary to disentangle the various mechanisms that
may cause decoherence but are not useful as a daily tool for analyzing experiments.
Therefore, it is of interest to examine the possibility whether a simple
phenomenological model can capture the essence of the physics of the full microscopic model.
Based on our results, presented in Section~\ref{results}, we propose to use a single-spin
model to which we artificially add a dephasing/relaxation mechanism.

Specifically, we propose that the Heisenberg equation of motion (in the rotating frame) of the expectation values
of the spin-components is modified according to
\begin{eqnarray}
\frac{\partial}{\partial t}
\langle \mathbf{S}(t) \rangle
&=&
\begin{pmatrix}
-1/T_2&2\pi\xi^z F_0&0\\
-2\pi\xi^z F_0&-1/T_2&\pi h_p(1+\zeta) (2+\xi^x+\xi^y)F_\mathrm{R}\\
0&-\pi h_p(1+\zeta)(2+\xi^x+\xi^y) F_\mathrm{R}& -1/T_1
\end{pmatrix}
\langle \mathbf{S}(t) \rangle
,
\label{BE0}
\end{eqnarray}
where we adopt the same notation as the one used in Section~\ref{modelparameter}.
The phenomenological aspect enters in the introduction of the decay times $T_1$ and $T_2$.

Equation~(\ref{BE0}) has the same structure as the Bloch equation  but there is a conceptual difference
and a practical consequence.
The former comes from the introduction of $g$-factor and microwave field amplitude distributions and the latter offers the
possibility to calculate numerically the effects of one-spin decoherence to a high degree of accuracy.
As we showed in this paper, one-spin decoherence plays an essential role when several qubits act at the same time.
It is then natural to start from the well-known equation of motion of a spin $S=1/2$,
add disorder through distribution probabilities (here of $g$-factors and
microwave field amplitude) and average over the solutions.
This leads to the exact knowledge of corresponding one-spin decoherence, namely
to Eq.~(\ref{BE0}) without the $T_1$ and $T_2$ terms.
If we now want to make a link with the Bloch equations we have just to add
the phenomenological damping times $T_1$ and $T_2$
as it is done in the original Bloch equations.
The difference between Eq.~(\ref{BE0}) and the original Bloch equations
is that in the latter $T_1$ and $T_2$ include all damping contributions i.e. many-spin
and one-spin damping, whereas in the former $T_1$ and $T_2$ include many-spins damping only,
one-spin damping being calculated exactly.

Before assessing the usefulness of Eq.~(\ref{BE0}) by comparing its results to the numerical
solution of the TDSE of the interacting spin system, it
is instructive to analyze the case $\xi^x=\xi^y=\xi^z=\zeta=T_1=0$.
Then the solution of Eq.~(\ref{BE0}) reads
\begin{eqnarray}
\langle \mathbf{S}^x(t) \rangle
&=&e^{-t/T_2}
\langle \mathbf{S}^x(0) \rangle
\nonumber \\
\langle \mathbf{S}^z(t) \rangle
&=&e^{-t/2T_2}\cos \left( 2\pi h_pF_\mathrm{R}\sqrt{1-(1/4\pi h_pF_\mathrm{R}T_2)^2}\right)
\langle \mathbf{S}^z(0) \rangle
,
\label{BE1}
\end{eqnarray}
where, for simplicity, we have assumed that $\langle \mathbf{S}^y(0) \rangle=0$.
From Eq.~(\ref{BE1}) it follows that the transverse and longitudinal
magnetization decays exponentially with a relaxation time $T_2$ and $2T_2$, respectively.
In other words, in the absence of randomness and for $T_1=0$, Eq.~(\ref{BE0}) predicts
a factor of two between the relaxation time of the Rabi oscillations
and the relaxation time of the transverse magnetization,
in qualitative (and almost quantitative) agreement with our simulation results
of dipolar-coupled spin-1/2 systems with randomness.
Thus, model Eq.~(\ref{BE0}) may give a simple explanation why in our simulations, we find
that extrapolation of $c_R$ to $h_p=0$ gives, in the presence of dipolar interactions, precisely $c_2/2$
if there is no distribution of $g$-factors ($D_0\not=0$, $\Gamma=0$) and a value larger than $c_2$
if there is a distribution of $g$-factors ($\Gamma>0$).

If we put $\xi^x=\xi^y=\xi^z=\zeta=0$, which in principle we should do if we strictly adopt the Bloch-equations approach,
we can never recover the linear dependence of the decay rate $1/T_R$ on the microwave amplitude $h_p$.
However, if we average over the $\xi$'s and/or $\zeta$ and put $T_2=\infty$, the results
are the same as those obtained from the direct solution of the TDSE of the spin-1/2 system.

\begin{figure}[t]
\includegraphics[width=8cm]{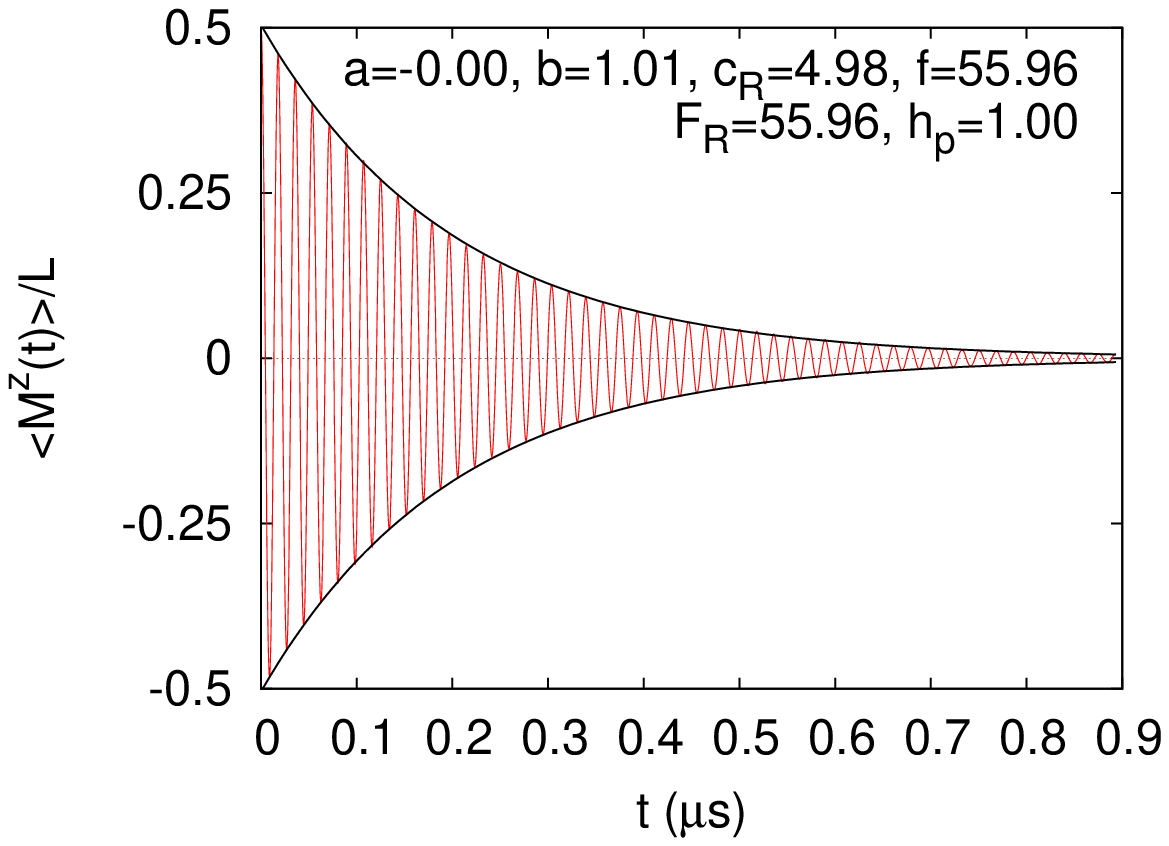}
\includegraphics[width=8cm]{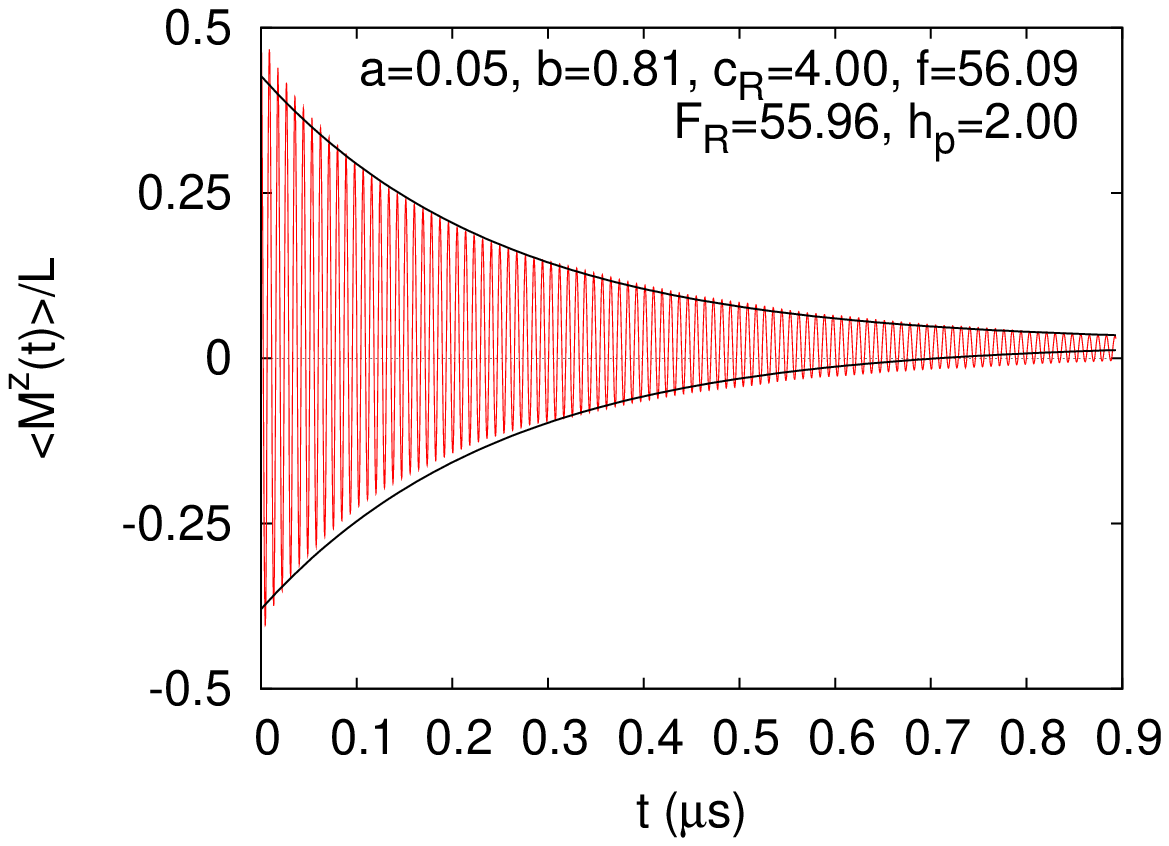}
\caption{(Color online)
The Rabi oscillations of the longitudinal magnetization as obtained from the numerical solution
of the phenomenological model Eq.~({\ref{BE0}}) with $T_1=\infty$, using 10000 realizations
of the random variables $\xi^x$, $\xi^y$, $\xi^z$, and $\zeta$.
Left: $\Gamma=0$, $\gamma=0.01$, $1/T_1=0$ and $T_2=3.0$, compare with Fig.~\ref{fig.Bb}(top right);
Right: $\Gamma=0.001$, $\gamma=0$, $1/T_1=0$ and $T_2=3.6$, compare with Fig.~\ref{fig.Cb.1}(middle right).
The solid line represents the envelope
$(a\pm be^{-c_\mathrm{R}t})/2$ of the function $(a+be^{-c_\mathrm{R}t}\cos2\pi ft)/2$ that was fitted
to the data.
}
\label{fig.BE}
\end{figure}

In appendix \ref{BE.A} we give a simple, robust, unconditionally stable algorithm~\cite{RAED87} to solve Eq.~(\ref{BE0}).
In Fig.~\ref{fig.BE} we present some representative results.
We used the same parameters for $\Gamma$, $\gamma$ and $h_p$ and changed the phenomenological
parameter $T_2$ until we found a fair match with the data of the corresponding interacting system.
Taking into account that we did not attempt to make a best fit to these data,
the agreement is excellent.
In both cases shown in Fig.~\ref{fig.BE} (and in many others cases not shown), this simple procedure seems to work quite well.
This suggests that the simple model Eq.~(\ref{BE0}) may be very useful for the analysis of experimental data,
including the effects of the pulse sequence and pulse shapes, effects that are rather expensive
to analyze using the large-scale simulation approach adopted in the present paper.

\section{Summary and outlook}\label{conclusions}

The main results of this paper may be summarized as follows:
\begin{itemize}
\item{The non-interacting spin model can account for the $\Omega_R$-dependence of the decay
of the Rabi oscillations if we introduce randomness
in the $g$-factors (all three) and/or in the amplitude of the microwave field.
In the case of $g_z$ randomness, the long-time average of the longitudinal magnetization deviates from zero.
This deviation increases as the Rabi frequency decreases and
reaches its maximum (1/2) when $h_p F_R/\Gamma F_0\rightarrow0$.
The effect of the $g_z$ distribution on the value of $c_R$ at zero microwave field ($h_p=0$) is
simply related to the value of $\langle M^z (t=\infty)\rangle$, suggesting that this decoherence effect comes from the
combination of different spin precessions about the $z$-axes and the nutational motion of spins.
}
\item{The dipolar-coupled spin system without randomness
in all three $g$-factors and without randomness in the amplitude of the microwave field,
cannot account for the $\Omega_R$-dependence of the Rabi oscillation decay rate, observed in experiment.
The decay rate of the Rabi oscillations increases as the concentration
of magnetic moments increases, as one naively would expect.
}
\item{The dipolar-coupled spin system without randomness
in $g_z$ but with randomness in the amplitude of the microwave field
and/or randomness in $(g_x,g_y)$,
can account for the $h_p$-dependence of the Rabi oscillation decay rate
and also for the concentration-dependence of this decay rate, just
as in the case of non-interacting spins.
}
\item{The dipolar-coupled spin system with randomness
in all three the $g$-factors and with or without randomness in the amplitude of the microwave field,
can account for the $h_p$-dependence of the Rabi oscillation decay rate
and also for the concentration-dependence of this decay rate.
A salient feature of the presence of fluctuations on $g_z$ (or, equivalently
on inhomogeneities in the static field) is that
the long-time average of the longitudinal magnetization deviates from zero,
as in the case of non-interacting spins.
}
\end{itemize}

For future work, we want to mention that the effects on the decay of the Rabi oscillations
of the measurement by the spin-echo pulses themselves may be studied
by the simple phenomenological model described in Section~\ref{BE}.
Among other aspects, not touched upon in the present study, are the case where
motional narrowing is important~\cite{ANDE53}
or where dipolar interactions are strong enough to induce decoherence by magnons,
as recently shown in the Fe$_8$ single molecular magnet~\cite{Takahashi11}.
These cases can be treated by the simulation approach adopted in this paper
and we plan to report on the results of such simulations in the near future.

\section*{Acknowledgements}
This work is supported by NCF, The Netherlands (HDR),
the Mitsubishi Foundation (SM) and
the city of Marseille, Aix-Marseille University (SB, BQR grant).
We thank the multidisciplinary EPR facility of Marseille (PFM Saint Charles) for technical support.

\appendix
\section{Overview of the model parameters}\label{parameters}

For convenience, we list the parameters of our model:
\begin{itemize}
\item{The Larmor frequency $F_0=\omega_0/2\pi\hbar=9700\,[\mathrm{MHz}]$ which is fixed.}
\item{The Rabi frequency at a microwave amplitude of 1 mT is $F_{\mathrm{R}}=55.96\,[\mathrm{MHz}]$ which is fixed.
}
\item{The amplitude of the microwave pulse, controlled by the parameter $h_p$. By convention,
if $h_p=1$, a single isolated spin will perform Rabi oscillations with a frequency of $F_{\mathrm{R}}=55.96\,[\mathrm{MHz}]$.
The Rabi pulsation in the microwave field $h_p$ is $\Omega_R=2\pi F_R h_p$.
}
\item{The width $\gamma$ of the Lorentzian distribution of the random fluctuations of the
amplitude of the microwave pulse $h_p$.}
\item{The width $\Gamma$ of the Lorentzian distribution of the random fluctuations of $g_x$, $g_y$, and $g_z$.
Unless mentioned explicitly, we assume that $g_x$, $g_y$, and $g_z$ share the same distribution.}
\item{The dipole-dipole coupling strength $D_0=51.88\,\mathrm{GHz}$, which is fixed.}
\item{The concentration $\concentration$ of magnetic impurities on the diamond lattice.}
\end{itemize}
\section{Numerical solution of the phenomenological model}\label{BE.A}
As in the case of the Bloch equations, if the relaxation time $T_1$ is finite, it is
useful to be able to specify both the initial value $\langle \mathbf{S}(t=0) \rangle=\langle \mathbf{S}(0) \rangle$
of the magnetization and its stationary-state  value $\langle \mathbf{S}(t=\infty) \rangle=\langle \mathbf{S}\rangle_{0}$.
Therefore, we extend Eq.~(\ref{BE0}) to
\begin{eqnarray}
\frac{\partial}{\partial t}
\langle \mathbf{S}(t) \rangle
&=&
\mathbf{A}\langle \mathbf{S}(t) \rangle+\mathbf{b}
,
\label{BE2}
\end{eqnarray}
where
\begin{eqnarray}
\mathbf{A}
&=&
\begin{pmatrix}
-1/T_2&2\pi\xi^z F_0&0\\
-2\pi\xi^z F_0&-1/T_2&\pi h_p(1+\zeta)(2+\xi^x+\xi^y)F_\mathrm{R}\\
0&-\pi h_p(1+\zeta)(2+\xi^x+\xi^y) F_\mathrm{R}& -1/T_1
\end{pmatrix}
,
\label{BE3}
\end{eqnarray}
and $\mathbf{b}^T=\langle \mathbf{S} \rangle_{0}/T_1$.
The formal solution of Eq.~(\ref{BE2}) reads
\begin{eqnarray}
\mathbf{\langle \mathbf{S}(t) \rangle}(t)&=&e^{{t }\mathbf{A}}\langle \mathbf{S}(0) \rangle+\int_0^t e^{u\mathbf{A}}\mathbf{b}\, du
\nonumber \\
&=&e^{t \mathbf{A}}\langle \mathbf{S}(0) \rangle+\mathbf{A}^{-1}(1-e^{t\mathbf{A}})\mathbf{b}
.
\label{BE4}
\end{eqnarray}
We integrate Eq.~(\ref{BE2}), that is we compute $e^{t \mathbf{A}}$, using the product-formula
$e^{t \mathbf{A}}=\left( e^{\tau \mathbf{A}_1/2}e^{\tau \mathbf{A}_2}e^{\tau \mathbf{A}_1/2}\right)^m + {\cal O}(\tau^3)$~\cite{SUZU85}
where $\tau=t/m$, $\mathbf{A}=\mathbf{A}_1+\mathbf{A}_2$ and
\begin{eqnarray}
\mathbf{A}_1
&=&
\begin{pmatrix}
-1/T_2&0&0\\
0&-1/T_2&0\\
0&0&-1/T_1
\end{pmatrix}
\nonumber \\
\mathbf{A}_2
&=&
\begin{pmatrix}
0&2\pi\xi^z F_0&0\\
-2\pi\xi^z F_0&0&\pi h_p(1+\zeta)(2+\xi^x+\xi^y) F_\mathrm{R}\\
0&-\pi h_p(1+\zeta)(2+\xi^x+\xi^y) F_\mathrm{R}&0
\end{pmatrix}
.
\label{BE7}
\end{eqnarray}
In detail, we have
\begin{eqnarray}
e^{\tau \mathbf{A}_1}&=&
\begin{pmatrix}
e^{-1/T_2}&0&0\\
0&e^{-1/T_2}&0\\
0&0&e^{-1/T_1}
\end{pmatrix}
\nonumber \\
e^{\tau \mathbf{A}_2}&=&
\begin{pmatrix}
1-(b/\Omega)^2 (1-\cos\tau \Omega) &  (b/\Omega) \sin\tau \Omega & (ab/\Omega^2)(1-\cos\tau \Omega) \\
-(b/\Omega) \sin\tau \Omega & \cos\tau \Omega & (a/\Omega) \sin\tau \Omega \\
(ab/\Omega^2)(1-\cos\tau \Omega) & -(a/\Omega) \sin\tau \Omega &1-(a/\Omega)^2 (1-\cos\tau \Omega)
\end{pmatrix}
,
\label{BE8}
\end{eqnarray}
where $a=2\pi\xi^z F_0$, $b=\pi h_p (1+\zeta)(1+\xi^x+\xi^y)F_\mathrm{R}$, and $\Omega=(a^2+b^2)^{1/2}$.

\bibliography{sylvain/biblio,c:/d/papers/epr11}   

\end{document}